\documentclass[12pt]{article}
\usepackage{amsmath}
\usepackage{bm}
\usepackage{graphicx}
\usepackage{enumerate}
\usepackage{natbib}
\usepackage{hyperref}
\usepackage{url} 
\usepackage{algorithm,algpseudocode} 
\usepackage{forest} 
\usepackage{booktabs} 
\usepackage{lscape}
\usepackage{longtable}
\usepackage{multirow, array}
\usepackage{threeparttable}
\usepackage{float}

\hypersetup{
    colorlinks=true,
    linkcolor=blue,
    citecolor=blue,
    filecolor=magenta,      
    urlcolor=blue,
    pdfpagemode=FullScreen,
    }

\newcommand{\balpha}{{\bm \alpha}}
\newcommand{\bbeta}{{\bm \beta}}

\newcommand{\bSigma}{\bm{\Sigma}}

\newcommand{\bV}{\mathbf{V}}

\newcommand{\bp}{\mathbf{p}}

\newcommand{\bv}{\mathbf{v}}
\newcommand{\bw}{\mathbf{w}}
\newcommand{\bx}{\mathbf{x}}
\newcommand{\by}{\mathbf{y}}
\newcommand{\bz}{\mathbf{z}}

\newcommand{\cC}{\mathcal{C}}

\newcommand{\cL}{\mathcal{L}}
\newcommand{\cM}{\mathcal{M}}

\newcommand{\cT}{\mathcal{T}}

\newcommand{\cW}{\mathcal{W}}

\newcommand{\bzero}{\bm{0}}

\newcommand{\blind}{1}

\begin{document}

\def\spacingset#1{\renewcommand{\baselinestretch}%
{#1}\small\normalsize} \spacingset{1}


\if1\blind
{
  \title{\bf Estimating Heterogeneous Exposure Effects in the Case-Crossover Design using BART}
  \author{Jacob R. Englert\\
    Department of Biostatistics and Bioinformatics, Emory University\\[1ex]
    Stefanie T. Ebelt \\
    Department of Environmental Health, Emory University \\[1ex]
    Howard H. Chang \\
    Department of Biostatistics and Bioinformatics, Emory University}
  \maketitle
} \fi

\if0\blind
{
  \bigskip
  \bigskip
  \bigskip
  \begin{center}
    {\LARGE\bf Estimating Heterogeneous Exposure Effects in the Case-Crossover Design using BART}
\end{center}
  \medskip
} \fi

\bigskip
\begin{abstract}
Epidemiological approaches for examining human health responses to environmental exposures in observational studies often control for confounding by implementing clever matching schemes and using statistical methods based on conditional likelihood. Nonparametric regression models have surged in popularity in recent years as a tool for estimating individual-level heterogeneous effects, which provide a more detailed picture of the exposure-response relationship but can also be aggregated to obtain improved marginal estimates at the population level. In this work we incorporate Bayesian additive regression trees (BART) into the conditional logistic regression model to identify heterogeneous exposure effects in a case-crossover design. Conditional logistic BART (CL-BART) utilizes reversible jump Markov chain Monte Carlo to bypass the conditional conjugacy requirement of the original BART algorithm. Our work is motivated by the growing interest in identifying subpopulations more vulnerable to environmental exposures. We apply CL-BART to a study of the impact of heat waves on people with Alzheimer's disease in California and effect modification by other chronic conditions. Through this application, we also describe strategies to examine heterogeneous odds ratios through variable importance, partial dependence, and lower-dimensional summaries.
\end{abstract}

\noindent%
\vfill

\newpage
\spacingset{1.9} 

\section{Introduction}
\label{sec:intro}

In the United States, Alzheimer's disease (AD) affects 6.7 million people aged 65 and older in 2023, with that number projected to more than double by 2060. AD is the most common cause of dementia, entirely or partially responsible for 60-80\% of all cases. People with AD often struggle to communicate and complete tasks in their daily life due to a host of symptoms headlined by forgetfulness, lethargy, and confusion. An estimated 1.3\% of emergency department (ED) visits involve people with AD and related dementia, and within this population the number of ED visits per 1,000 Medicare beneficiaries increased 28\% from 2008 to 2018 - outpacing cancer, stroke, and heart failure \citep{alzheimers_association_2023_2023}.

In recent years, extreme heat has been associated with elevated risk of ED visit, hospitalization, and death among people with AD and dementia in Spain, Australia, Germany, and the United States \citep{linares_short-term_2017, culqui_association_2017, xu_heatwaves_2019, fritze_effect_2020, wei_associations_2019, zhang_short-term_2023}. \cite{fritze_effect_2020} reported that the number of comorbid conditions is associated with increased risk of mortality among people with dementia.

There are several potential explanations for why people with AD are more affected by extreme heat. People with AD may have elevated core body temperature due to disturbed circadian rhythms responsible for thermoregulation \citep{volicer_sundowning_2001, satlin_circadian_1995, harper_disturbance_2005, klegeris_increase_2007}. Alternatively, people with AD tend to wander or get lost, resulting in prolonged exposure to extreme temperatures \citep{alzheimers_association_2023_2023}. Another possibility is that these individuals may struggle to communicate their heat-related discomfort in certain situations with their caregivers \citep{van_hoof_thermal_2010}.

These explanations may not apply to the entire AD population. People with AD and related dementia are 2.7 times more likely to have 4 or more additional chronic conditions compared to people without AD or dementia; in the United States, 56\% have hypertension, 46\% have chronic kidney disease (CKD), 37\% have diabetes, 34\% have congestive heart failure (CHF), and 20\% have chronic obstructive pulmonary disease (COPD) \citep{alzheimers_association_2023_2023}. Additionally, an estimated 12.7\% of people with AD have depression \citep{chi_prevalence_2015}. Given the variety of concomitant diagnoses these individuals tend to have, it is possible that heterogeneity exists in the exposure-response relationship describing heat wave-related morbidity. Thus, studying modifiers of this relationship is of interest.

Some of the aforementioned studies model heterogeneous exposure effects via stratification or by interacting covariates with the exposure \citep{fritze_effect_2020, zhang_short-term_2023}, while others directly target these effects using the case-only approach of \citet{armstrong_fixed_2003} \citep{xu_heatwaves_2019}. In the extreme temperature literature, stratification has been mostly applied for demographic characteristics (e.g., age, race, and sex), while the case-only approach has also been used for chronic conditions, socioeconomic status, and various census tract characteristics \citep{schwartz_who_2005, zanobetti_susceptibility_2013, xu_heatwave_2017, madrigano_case-only_2015}. These approaches to heterogeneous effects estimation are limited by the need for expert knowledge regarding which factors are important before model-fitting, and they are not readily capable of identifying complex interactions among potential effect moderators.

We propose an extension of the popular case-crossover study design to estimate heterogeneous exposure effects using Bayesian additive regression trees (BART) \citep{chipman_bart_2010}. As it is typically applied, this design is limited to the previously mentioned strategies for heterogeneous effects estimation. The proposed method, CL-BART, uses BART within the case-crossover design to flexibly learn potentially complicated heterogeneous exposure-response relationships during the model-fitting process, with minimal prespecification required. In Section \ref{sec:data} we introduce the data for the motivating application. In Section \ref{sec:methods} we review the case-crossover design, conditional logistic regression, and BART. We then develop CL-BART, focusing on the reversible jump portion of the estimation algorithm. In Section \ref{sec:inference} we describe strategies for drawing posterior inference from the proposed model. In Section \ref{sec:simulation}, we conduct two simulations illustrating the performance of CL-BART, and in Section \ref{sec:application} we apply CL-BART to estimate the effects of heat waves on ED visits among people with AD in California. Finally, in Section \ref{sec:discussion} we summarize our findings, discuss the limitations of the approach, and suggest possibilities for future improvements.

\section{Data}
\label{sec:data}
\subsection{Health Data}
The data for our motivating application includes all ED visits among people with AD in California occurring from 2005 to 2015. These data were obtained from the California Office of Statewide Health Planning and Development, and include patients' visit date, sex, age, race, ethnicity, residential ZIP code, and diagnosis codes based on the International Classification of Diseases. We restrict the ED visit records to include only those who had either a primary or secondary diagnosis of AD. Diagnoses of comorbid conditions were also based on the presence of any diagnosis code for CHF, CKD, COPD, depression, diabetes, hypertension, and hyperlipidemia (see the Supplementary Materials for a list of codes).

\subsection{Exposure Data}
Meteorology data were obtained from Daymet \citep{thornton_daymet_2022}. The 1 km data product was spatially averaged within each ZIP code, and linked to the ED visit data by both date and ZIP code. Specifically, we use the daily average temperature ($^{\circ}$C) and dew-point temperature ($^{\circ}$C). The former is calculated as the arithmetic mean of the daily minimum and maximum temperature, and the latter is derived from water vapor pressure using the Magnus formula presented in \citet{sonntag_important_1990}. The exposure of interest, heat wave, is defined as any sequence of two or more days at or above the ZIP code-specific 95th percentile of daily average temperature (excluding the first day of such a sequence to better reflect sustained heat exposure). Daily average temperature, dew-point temperature, and a US federal holiday indicator were also treated as potential confounders in the health model.

\section{Methods}
\label{sec:methods}

\subsection{Model Specification}

\subsubsection{Case-Crossover Design and Conditional Logistic Regression}
Case-crossover designs are frequently used to analyze the effects of short-term exposure on health outcomes in environmental epidemiology studies \citep{carracedo-martinez_case-crossover_2010}, when only cases are available. In the case-crossover design, cases are matched to a set of controls within a ``referent window" to create a stratum. A popular strategy for selecting this window is the time-stratified approach, which matches each case to the 3-4 other dates in a calendar month with the same day of the week. This strategy assumes the cases are independent and rare enough such that an individual would not experience the event twice within the referent window. The time-stratified design is both localizable and ignorable, thus providing unbiased estimation of regression coefficients when using conditional logistic regression \citep{janes_case-crossover_2005}.

Suppose $n$ cases are observed. Then the true data generating model is:
\begin{gather}
\label{true-dgp}
Y_{it} \sim \text{Bernoulli}(p_{it}), \qquad i = 1,\ldots,n, \qquad t\in\cW_i \\
\label{logit-link}
\text{logit} (p_{it}) =  \bv_i^T\balpha + \bx_{it}^T\bbeta + \tau z_{it},
\end{gather}
where $Y_{it}$ is the outcome for individual $i$ at time $t$, $p_{it}$ is the probability of observing $Y_{it} = 1$, and $\cW_i$ is the referent window containing observation times $t$ for individual $i$. The primary exposure is denoted by $z_{it}$, while $\bv_i$ and $\bx_{it}$ represent column vectors which include time-invariant and time-varying confounders of the exposure-response relationship, respectively. For the AD example, $z_{it}$ is a binary heat wave indicator, $\bv_i$ includes time-invariant demographic information or other unmeasured quantities, and $\bx_{it}$ includes daily average temperature, dew-point temperature, and a federal holiday indicator.

Operating under the case-crossover assumption that $\sum_{t \in \cW_i} Y_{it} = 1, \quad \forall$ $i = 1, \ldots, n$, the conditional likelihood for the observed data is given by \eqref{clr-lik}
\begin{equation}
\label{clr-lik}
    \pi(\by \mid \tau, \bbeta) = \prod_{i=1}^n \pi(\by_i \mid \tau, \bbeta) = \prod_{i=1}^n \frac{\exp \left\{\bx_{it_i}^T\bbeta + \tau z_{it_i} \right\}}{\sum_{t \in \cW_i} \exp \left\{\bx_{it}^T\bbeta + \tau z_{it} \right\}},
\end{equation}
where $\by_i$ is the observed vector of outcomes within referent window $\cW_i$, and $\by$ is a vector containing all $\by_i$, $i = 1, \ldots, n$. Notably, $\by_i$ is only implicitly present through the subscript $it_i$, which represents the index time point of the $i^{th}$ case, and all time-invariant covariates ($\bv_i$) have been conditioned out entirely. Maximum likelihood estimation of \eqref{clr-lik} results in unbiased log odds ratios (ORs) for the confounders ($\bbeta$) and the primary exposure ($\tau$).

This model assumes a homogeneous exposure effect across individuals. To examine how the association between the exposure and outcome varies across individuals, researchers may specify subgroup analyses ahead of time, defining the subgroups using demographic characteristics like sex and age. This requires some knowledge of the outcome and exposure to be able to identify which subgroups should be considered.

\subsubsection[Heterogeneous Exposure Effects with CL-BART]{CL-BART and the Exposure Moderating Function $\tau(\cdot)$}
We propose extending this framework to allow for estimation of heterogeneous exposure effects within a study population. Specifically, we use BART \citep{chipman_bart_2010} to model the exposure effect as a function of individual-level covariates that were previously conditioned out. We start by defining a more general version of the conditional likelihood in \eqref{clbart-lik}, which we will refer to as the conditional logistic BART (CL-BART) likelihood.
\begin{equation}
\label{clbart-lik}
    \pi(\by \mid \tau(\cdot), \bbeta) = \prod_{i=1}^n \pi(\by_i \mid \tau(\bw_i), \bbeta) = \prod_{i=1}^n \frac{\exp \left\{\bx_{it_i}^T\bbeta + \tau(\bw_i) z_{it_i} \right\}}{\sum_{t \in \cW_i} \exp \left\{\bx_{it}^T\bbeta + \tau(\bw_i) z_{it} \right\}}.
\end{equation}

Here we have simply replaced $\tau$ with $\tau(\bw_i)$, suggesting that the increase in the log odds of an ED visit due to a unit increase in $z$ may differ across individuals. The contents of $\bw_i$ may overlap with $\bv_i$ in \eqref{logit-link}, but the two need not be identical. In the AD example, $\bw_i$ includes comorbid conditions, such as diabetes and CKD, as well as sex and age.

We place a nonparametric BART prior on the exposure moderating function as in \eqref{tau}.
\begin{equation}
\label{tau}
\tau(\bw_i) = \sum_{m = 1}^{M} g(\bw_i \mid \cT_m, \cM_m).
\end{equation}
The BART prior represents $\tau(\bw_i)$ as a sum of $M$ weak learners - in this case, Bayesian regression trees \citep{chipman_bayesian_1998}. Each tree is composed of a tree structure $\cT$ defined by a series of binary splits based on covariates $\bw$, a set of terminal or ``leaf" nodes $\cL(\cT)$, and a set of scalar leaf node parameters $\cM = \left\{\mu_l \right\}_{l \in \cL(\cT)}$. In \eqref{tau}, $g$ is the function that makes a prediction for covariates $\bw$ by mapping $\bw$ to a single leaf node in the given tree.

In the simplest setting, $\bw$ consists only of a series of $P$ binary effect moderators and the maximum number of unique values of $\tau(\bw)$ is $2^P$, regardless of sample size. When $M = 1$, CL-BART simplifies to a treed conditional logistic regression, where the confounder effects are shared across leaf nodes. Including continuous covariates is a straightforward extension, and allows for modeling more complex high-order interactions and nonlinearities among exposure effect moderators. We do not consider time-varying covariates in $\bw$, as this would result in individual strata being allocated to multiple leaf nodes, thus violating the case-crossover design.

\subsection{Estimation}
\subsubsection{Generalized BART}
BART was originally designed with Gaussian outcomes in mind, relying heavily on the conditional conjugacy between the outcome model and the prior distribution on the leaf node parameters. This allows for a Metropolis-Hastings (M-H) proposal for the tree structure to be conducted separately from the Gibbs update of the leaf node parameters through marginalization, resulting in a simple and efficient Markov chain Monte Carlo (MCMC) algorithm \citep{chipman_bart_2010}. BART has since been extended to other outcome regressions, including survival, log-linear, and gamma models \citep{sparapani_nonparametric_2016, murray_log-linear_2021, linero_semiparametric_2020}, but such extensions require extensive modification of the original algorithm. BART has also been used to model varying coefficients \citep{deshpande_vcbart_2023, hahn_bayesian_2020}, but these leverage conditional conjugacy as well. Recently, \cite{linero_generalized_2024} proposed a general strategy based on reversible-jump MCMC (RJMCMC) \citep{green_reversible_1995, godsill_relationship_2001} as a promising alternative for adapting BART to more complicated likelihoods. This approach is appealing because it avoids the need for conjugate priors altogether, and so we use it for CL-BART. We now provide a brief overview of this approach and our implementation, but refer the reader to the source for further detail.

\subsubsection{Data Likelihood}
It is first helpful to rewrite the data likelihood in terms of the tree structure. The likelihood for a single tree $\cT_m$ can be represented as in \eqref{tree-lik}.
\begin{equation}
\label{tree-lik}
    \pi(\by \mid \cT_m, \cM_m) = \prod_{l \in \cL(\cT_m)} \prod_{i:\bw_i \mapsto l} \pi \left(\by_i \mid \mu_l \right).
\end{equation}

For CL-BART, we may substitute the likelihood given in \eqref{clr-lik}, where $\mu_l$ represents the prediction from $\cT_m$ (i.e., the exposure effect) for strata having $\bw_i$ mapped ($\mapsto$) to leaf node $l$. Since $\bbeta$ is shared across all leaf nodes, we omit it in \eqref{tree-lik} to lighten the notation.

\subsubsection{Prior Distribution}
\label{sec:priors}
The unknown quantities for each tree $m$ in CL-BART include the leaf node parameters $\cM_m = \left\{\mu_l\right\}_{l\in \cL(\cT_m)}$ and the tree structure itself $\cT_m$. By imposing independence on the former, we may factor the joint prior distribution for a single tree $m$ as in \eqref{tree-priors}.
\begin{equation}
\label{tree-priors}
    \pi\left(\cT_m, \cM_m \right) \sim \pi(\cT_m) \pi(\cM_m \mid \cT_m) \sim \pi(\cT_m) \prod_{l \in \cL(\cT_m)} \pi\left(\mu_l\right).
\end{equation}

The $\mu_l$ are given i.i.d. Normal($0$, $\sigma_\mu^2$) priors, but this is not a requirement since conjugacy with the likelihood is no longer a concern. While one may have some intuition regarding the range of values to expect for $\tau(\cdot)$, generally this will be unknown. For this reason, we follow \citet{linero_generalized_2024} and specify a half-Cauchy hyperprior $\sigma_\mu \sim \cC_+\left(0, k / \sqrt{M}\right)$ to help learn the range of appropriate predictions. Here, $k$ is a fixed hyperparameter, and the division by $\sqrt{M}$ ensures predictions are made on the same general scale regardless of the number of trees used.

For the tree structure $\cT_m$, we use the ``branching process" prior described in \cite{chipman_bart_2010}, where each node in $\cT_m$ is split with probability $\rho_d = \gamma(1 + d)^{-\xi}$ (here $d$ is the depth of the node in $\cT_m$). We use the default values of $(\gamma, \xi) = (0.95, 2)$, but note that in the heterogeneous effects setting there have been several proponents for stronger regularization \citep{hahn_bayesian_2020, caron_shrinkage_2022}. We make one departure from the traditional branching process by further placing a Dirichlet hyperprior on the covariate selection probabilities as suggested in \citet{linero_bayesian_2018}. This modification helps particularly in settings with many covariates that each have many available values upon which to split. 

\subsubsection{Tree Proposals and the Posterior Distribution}
\label{sec:tree-props}
New tree structures are proposed and accepted with a M-H step. We consider three types of proposals: grow, prune, and change. Both the grow and prune moves involve jumping between parameter spaces of differing dimensions, and thus require modification of the traditional M-H acceptance ratio. The general form for this ratio is given in \eqref{rjmcmc-mh-ratio}:
\begin{equation}
\label{rjmcmc-mh-ratio}
    r_{\cT} = \underbrace{\frac{\pi\left(\cT', \cM' \right)}{\pi\left( \cT, \cM \right)}}_{\text{Prior Ratio}} \times \underbrace{\frac{\pi\left(\by \mid \cT', \cM' \right)}{\pi\left(\by \mid \cT, \cM \right)}}_{\text{Likelihood Ratio}} \times \underbrace{\frac{\pi\left( \cT, \cM \mid \cT', \cM' \right)}{\pi\left( \cT', \cM' \mid \cT, \cM \right)}}_{\text{Proposal Ratio}}.
\end{equation}

The prior term may be factored as in Section \ref{sec:priors}, while the proposal term may be factored into two parts: a structural component and a proposal for the new leaf node parameter(s) based on some distribution $G$. We use a normal distribution based on a Laplace approximation for $G$, as suggested by \cite{linero_generalized_2024} (see the Supplementary Materials for details). Each type of proposal is summarized below, where $\text{NOG}(\cT)$ is defined as the set of nodes in $\cT$ that are parents of two terminal nodes.
\begin{itemize}
    \item \textbf{Grow}: a random node $l \in \cL(\cT)$ is selected. Subsequently, a splitting covariate $w_p$, and cut-point $C_l$ based on the values of $w_p$ are selected. Then $l$ is split into $lL$ and $lR$, where strata having $w_p \le C_l$ are fed into $lL$ and strata having $w_p > C_l$ are fed into $lR$. Assuming $l$ has depth $d$, the modified RJMCMC M-H acceptance ratio is:
    \begin{equation}
    \label{r_grow}
    \begin{split}
        r_{\cT}^{grow} & = \frac{\rho_d(1-\rho_{d+1})^2}{(1-\rho_d)} \times \frac{\pi(\mu_{lL}' \mid 0, \sigma_{\mu}^2) \times \pi(\mu_{lR}' \mid 0, \sigma_{\mu}^2)}{\pi(\mu_{l} \mid 0, \sigma_{\mu}^2)} \\[1ex]
        & \qquad \times \frac{\prod_{i:\bw_i \mapsto {lL}} \pi(\by_i \mid \mu_{lL}') \times \prod_{i:\bw_i \mapsto {lR}} \pi(\by_i \mid \mu_{lR}')}{\prod_{i:\bw_i \mapsto l} \pi(\by_i \mid \mu_l)} \\[1ex]
        & \qquad \times \frac{p_{prune}(\cT')|\text{NOG}(\cT')|^{-1}}{p_{grow}(\cT)|\cL(\cT)|^{-1}} \times \frac{G_{prune}(\mu_l)}{G_{grow}(\mu_{lL}', \mu_{lR}')}.
    \end{split}
    \end{equation}

    \item \textbf{Prune}: a random node $b \in \text{NOG}(\cT_m)$ is selected. Leaf nodes $bL$ and $bR$ are removed from the tree, along with the variable and cut-point that defined them. Assuming $b$ has depth $d$, the modified RJMCMC M-H acceptance ratio is:
    \begin{equation}
    \label{r_prune}
    \begin{split}
        r_{\cT}^{prune} & = \frac{(1-\rho_d)}{\rho_d(1-\rho_{d+1})^2} \times \frac{\pi(\mu_{l}' \mid 0, \sigma_{\mu}^2)}{\pi(\mu_{lL} \mid 0, \sigma_{\mu}^2) \times \pi(\mu_{lR} \mid 0, \sigma_{\mu}^2)}\\[1ex]
        & \qquad \times \frac{\prod_{i:\bw_i \mapsto b} \pi(\by_i \mid \mu_b')}{\prod_{i:\bw_i \mapsto {bL}} \pi(\by_i \mid \mu_{bL}) \times \prod_{i:\bw_i \mapsto {bR}} \pi(\by_i \mid \mu_{bR})} \\[1ex]
        & \qquad \times \frac{p_{grow}(\cT')|\cL(\cT')|^{-1}}{p_{prune}(\cT)|\text{NOG}(\cT)|^{-1}} \times \frac{G_{grow}(\mu_{bL}, \mu_{bR})}{G_{prune}(\mu_b')}.
    \end{split}
    \end{equation}

    \item \textbf{Change}: a random node $b \in \text{NOG}(\cT_m)$ is selected. The criteria for further splitting into leaf nodes $bL$ and $bR$ are exchanged for another variable and/or cut-point. Since the general tree structure is unchanged, the structural components of the prior and proposal ratios cancel out. The M-H acceptance ratio is:
    \begin{equation}
    \label{r_change}
    \begin{split}
    r_{\cT}^{change} & = \frac{\pi(\mu_{lL}' \mid 0, \sigma_{\mu}^2) \times \pi(\mu_{lR}' \mid 0, \sigma_{\mu}^2)}{\pi(\mu_{lL} \mid 0, \sigma_{\mu}^2) \times \pi(\mu_{lR} \mid 0, \sigma_{\mu}^2)} \\[1ex]
    & \qquad \times \frac{\prod_{i:\bw_i \mapsto {bL}} \pi(\by_i \mid \mu_{bL}') \times \prod_{i:\bw_i \mapsto {bR}} \pi(\by_i \mid \mu_{bR}')}{\prod_{i:\bw_i \mapsto {bL}} \pi(\by_i \mid \mu_{bL}) \times \prod_{i:\bw_i \mapsto {bR}} \pi(\by_i \mid \mu_{bR})} \\[1ex] 
    & \qquad \times \frac{G_{change}(\mu_{bL}, \mu_{bR})}{G_{change}(\mu_{bL}', \mu_{bR}')}.
    \end{split}
    \end{equation}

\end{itemize}

At each iteration, one type of proposal is made for each tree in the ensemble. We set the prior probability of each proposal type to $p_{grow}=0.3$, $p_{prune}=0.3$, and $p_{change}=0.4$. The trees are cycled through using a generalized version of Bayesian backfitting \citep{hastie_bayesian_2000, linero_generalized_2024}. Essentially, this involves offsetting the likelihood calculation in the M-H acceptance ratio for the update of tree $\cT_m$ by the sum of the predictions from the remaining $M-1$ trees. Mathematically, we swap \eqref{tree-lik} with \eqref{backfit-tree-lik}
\begin{equation}
\label{backfit-tree-lik}
    \pi(\by \mid \cT_m, \cM_m) = \prod_{l \in \cL(\cT_m)} \prod_{i:\bw_i \mapsto l} \pi\left(\by_i \mid \mu_l + \lambda_i^r \right),
\end{equation}
where $\lambda_i^r = \sum_{k \ne m} g(\bw_i \mid \cT_k, \cM_k)$. The M-H acceptance ratios presented in this section only depend on the likelihood within the affected leaf nodes, and so the inner product term of \eqref{backfit-tree-lik} can be used wherever the likelihood is evaluated in \eqref{r_grow}, \eqref{r_prune}, and \eqref{r_change}.

Thus far for tree $m$, the proposed values $\cM_m$ have been used solely to update $\cT_m$. Once $\cT_m$ has been updated, we propose new values for all $\mu_l \in \cM_m$ sequentially from their full conditional distribution via adaptive rejection sampling \citep{gilks_adaptive_1992}.

Prior to the BART update, we update $\bbeta$ using a traditional random-walk M-H step. We use a multivariate normal proposal distribution which is centered at the current value of $\bbeta$ and has covariance matrix $\sigma_{\beta}^2 \bV_\bbeta$, where we initialize $\bV_\bbeta$ as the confounder portion of the covariance matrix of $\hat{\bbeta}$ from the fit of a conventional conditional logistic regression as in \eqref{clr-lik}, and $\sigma_{\beta}^2$ is initially set to unity but tuned throughout the burn-in phase to achieve an optimal acceptance rate. Note that the proposal for $\left\{\cT_m, \cM_m \right\}$ is also offset by the confounders, in addition to the fits of other $M-1$ trees. An outline for the CL-BART MCMC algorithm is given in Algorithm \ref{alg:one}.

\spacingset{1.1}
\begin{algorithm}[t]
\caption{One MCMC iteration of CL-BART}\label{alg:one}
\begin{algorithmic}[1]

\State \textbf{Input:} $\text{Data}:\left\{\bw, \bx, \by, \bz\right\}, \bbeta, \left\{\cT_m,\cM_m \right\}_{m=1}^M, \gamma, \xi, \left\{s_p \right\}_{p=1}^P, a, \sigma_\mu^2, \sigma_\beta^2, \bV_\beta$

\State Update $\bbeta$ (via M-H step with multivariate normal prior).
\State Set $\lambda_i \gets \sum_{m=1}^M g(\bw_i \mid \cT_m, \cM_m)$ for $i = 1, \ldots, n$.
\For{$m = 1$ to $M$}
    \State Set $\lambda_i^r \gets \lambda_i - g(\bw_i \mid \cT_m, \cM_m)$ for $i = 1, \ldots, n$.
    \State Propose $\cT_m'$ from $\cT_m$ using a grow, prune, or change step.
    \State Propose $\cM_m^*$ from $G$ based on a Laplace approximation \citep{linero_generalized_2024}.
    \State Compute $r_{\cT}$, the (modified) M-H acceptance ratio for $\left\{\cT_m',\cM_m^* \right\}$.
    \State Set $\cT_m \gets \cT_m'$ with probability $\min(1, r_{\cT})$.
    \State Update $\cM_m \mid \cT_m, \left\{\lambda_i^r\right\}_{i=1}^n$ using adaptive rejection sampling \citep{gilks_adaptive_1992}.
    \State Set $\lambda_i \gets \lambda_i^r + g(\bw_i \mid \cT_m, \cM_m)$ for $i = 1, \ldots, n$.
\EndFor
\State Update $\left\{s_p \right\}_{p=1}^P \sim \text{Dirichlet}(\frac{a}{P} + u_1, \ldots, \frac{a}{P} + u_P)$, where $u_p$ is the number of times the $p^{th}$ element of $\bw$ is split upon.
\State Update $a$ (via discrete step described in \citet{linero_bayesian_2018}).
\State Update $\sigma_\mu$ (via M-H step with half-Cauchy prior).

\end{algorithmic}
\end{algorithm}
\spacingset{1.9}

\section{Posterior Inference}
\label{sec:inference}

As with any Bayesian model, point estimates and posterior credible intervals may be obtained for the confounder coefficients and other scalar parameters. To summarize the estimated heterogeneous exposure effects, we introduce estimands similar to those presented in the BART for causal inference literature \citep{hill_bayesian_2011, hahn_bayesian_2020, woody_estimating_2020}, with the two main differences being that we are working on the log odds ratio scale, and that we do not claim that our estimates have causal interpretations.


Initially, we estimate the average conditional exposure effect for a unit increase in the exposure as $\bar{\tau} = \frac{1}{n}\sum_{i=1}^n \hat{\tau}(\bw_i)$. This may also be exponentiated if an OR interpretation is desired. Perhaps of greater interest are the individual conditional exposure effects $\tau(\bw_i)$, $i = 1, \ldots, n$. These are numerous, so it is helpful to have strategies for summarizing them. We can easily obtain point estimates and posterior credible intervals of $\tau(\bw)$ for any desired set of exposure modifiers $\bw$. However, the individual-level quantities can be noisy, and so it can be beneficial to instead report partial averages of conditional exposure effects, such as the partial dependence functions introduced in \citet{friedman_greedy_2001}.

Define $\bw = (w^1,w^2,\ldots,w^P)^T$ as the $P$-vector of potential effect moderators, $\bw^{[\bp]}$ as the $\bp^{th}$ component of $\bw$ being evaluated (may be multiple components), and $\bw^{[-\bp]}$ as all but the $\bp^{th}$ component of $\bw$. The corresponding observations made on individual $i$ are $\bw_i$, $\bw^{[\bp]}_i$, and $\bw^{[-\bp]}_i$, respectively. The partial average exposure effect is estimated as in \eqref{pacee}.
\begin{equation}
\label{pacee}
\bar{\tau}_{partial}(\bw^{[\bp]} = \bw^*) = \frac{1}{n} \sum_{i=1}^n \hat{\tau}(\bw_i^{[-\bp]}, \bw^*).
\end{equation}

One might select multiple settings of $\bw^*$ for comparison, where only a subset of $\bw$ need be included in $\bw^*$, and calculate \eqref{pacee} for each setting. The resulting estimates (or any function of the estimates) can be compared across the posterior distribution. The simplest case is to fix a single binary covariate in $\bw$, compute \eqref{pacee} for both levels of the covariate, and then calculate the difference in the partial averages. The corresponding estimate represents the marginal contribution of the selected covariate to the average exposure effect.

It may be difficult or computationally infeasible to perform an exhaustive comparison of all partial average exposure effects. To identify covariate values to fix during the partial averaging in \eqref{pacee}, we suggest creating a lower-dimensional summary using, say, a single decision tree (e.g., the CART algorithm of \citet{breiman_classification_2017}) as described by \citet{woody_model_2021}. This involves using some subset of the input covariates $\bw$ to ``predict" the posterior mean individual exposure effects. We can then compute \eqref{pacee} for the combinations of covariates leading to each leaf node in the resulting CART summary.

\subsection{Model Diagnostics}
\label{diagnostics}

Sometimes it is helpful to have a ``quick and dirty" way to establish variable importance. One option is to summarize the frequencies with which the BART portion of the model splits on each of the effect moderators \citep{chipman_bart_2010}. In general, we expect the model to favor splits on covariates which are essential to the true data generating process. As the size of the ensemble increases, however, spurious splits will be included. Imposing sparsity on the branching process prior \citep{linero_bayesian_2018} helps to alleviate this issue in many cases. Additionally, it is important to consider correlation between the covariates, and that there may be more than one path to a good model.

Since predictions are not available when using conditional logistic regression, cross-validation based on model selection criteria that involve the outcome (e.g., RMSE or similar) do not apply. However, models can still be evaluated using likelihood-based criteria. We suggest referencing the Widely-Applicable Information Criteria (WAIC), which approximates leave-one-out cross-validation \citep{gelman_understanding_2014}. The WAIC uses the entire posterior distribution and all of the available data to evaluate and penalize models, and can conveniently be computed during model-fitting. This metric is useful for comparing CL-BART models with different hyperparameter specifications, such as the number of trees. 

Lastly, it is essential to monitor the convergence of the posterior distribution. Since the RJMCMC algorithm performs both model selection and parameter estimation, posterior chains of individual exposure effects may not have well-mixed trace plots due to the possibility of jumping between different parameter spaces. For this reason we suggest monitoring trace plots for global parameters, such as $\bar{\tau}$, $\bbeta$, $\sigma_\mu^2$, and other quantities such as the log-likelihood or average number of nodes across trees. We did not find it was necessary to run multiple chains for the analyses described in the simulation study and application.

\section{Simulation Study}
\label{sec:simulation}
In this section we design a simulation to mimic the case-crossover design. We follow 10,000 individuals for three years, and generate their shared exposure time-series as in \eqref{exposure}.
\begin{equation}
\label{exposure}
    Z_t \sim \text{Normal}\left(\sin\left(\frac{2 \pi t \times 3}{1096}\right), 1\right), \quad t = 1, \ldots, 1096.
\end{equation}
Five time-varying confounders are generated as $x_1, \ldots, x_5 \overset{i.i.d.}{\sim} \text{Uniform(0, 1)}$, with odds ratios 0.5, 0.8, 1.0, 1.2, and 2.0. The probability of individual $i$ experiencing the event at time $t$ is calculated as $p_{it} = \text{expit}\left(\alpha + \bx_{it}^T \bbeta + \tau(\bw_i) z_t \right)$, where $\alpha = -8$ to ensure rare events, and the true $\tau(\bw_i)$ for each individual $i$ is specified under two deterministic scenarios:
\begin{enumerate}
    \item \textbf{CART}: 10 binary covariates are generated as $\left[w_1, w_2, \ldots, w_{10} \right]^T \sim \text{MVN}(\bzero, \bSigma)$, where $\bSigma$ has an AR-1 structure (i.e., $\bSigma_{j,j'} = 0.6^{|j - j'|}$). Three of these 10 covariates are randomly selected ($w_1'$, $w_2'$, $w_3'$) and $\tau(\bw_i)$ is given one of four values according to the tree diagram in Figure \ref{fig:cart-sim-diagram}.
\begin{figure}[h]
\caption{True Conditional Odds Ratios for CART Simulation}
\label{fig:cart-sim-diagram}
\begin{center}
\begin{forest} for tree={
    edge path={\noexpand\path[\forestoption{edge}] (\forestOve{\forestove{@parent}}{name}.parent anchor) -- +(0,-12pt)-| (\forestove{name}.child anchor)\forestoption{edge label};}
}
[
[$w_1' \leq 0$
[$w_2' \leq 0$ [0.8]] [$w_2' > 0$ [1.1]]]
[$w_1' > 0$
[$w_3' \leq 0$ [1.3]] [$w_3' > 0$ [1.5]]]
]
\end{forest}
\end{center}
\end{figure}
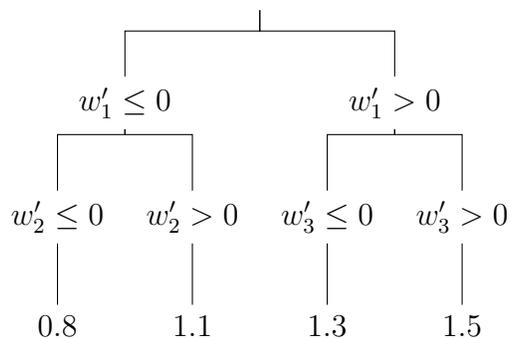
\vspace{-2em}
    \item \textbf{Friedman}: 10 continuous covariates are generated as $w_1, \ldots, w_{10} \overset{i.i.d.}{\sim} \text{Uniform(0, 1)}$, and $\tau(\bw_i) = [f(\bw_i) - 14] / 15$, where $f$ \eqref{friedman} is the benchmark function proposed in \citet{friedman_multivariate_1991}. We have scaled $f$ to approximately have a mean of zero and standard deviation of one-third, thus restricting the majority of potential odds ratios to be between 0.5 and 2.
    \begin{equation}
    \label{friedman}
    f(\bw) = 10 \sin(\pi w_1 w_2) + 20(w_3-0.5)^2 + 10w_4 + 5w_5
    \end{equation}
\end{enumerate}
As individuals are followed throughout the study period, cases are noted and the time-stratified case-crossover design is implemented. In both scenarios, approximately 4500 cases are typically observed.

For Scenario 1, we compare 1, 5, 10, 25, and 50 tree ensembles. For Scenario 2, due to the presence of many continuous predictors, we explore larger ensembles of 5, 10, 25, 50, and 100 trees. For both scenarios, we set $(k, \gamma, \xi) = (1, 0.95, 2)$ and run 10,000 total MCMC iterations, with the first 5,000 serving as a burn-in period. We keep every fifth post-burn-in sample, resulting in 1,000 posterior samples. Other hyperparameter settings are explored in the Supplementary Materials.

To evaluate performance we fit an ``oracle" conditional logistic regression by creating a design matrix consisting of the true interactions and/or functional forms of the moderators, each interacting with the exposure. For each simulation run we compute the average bias \eqref{avg-bias}, root mean square error (RMSE) \eqref{rmse}, and average 95\% posterior credible interval coverage \eqref{avg-coverage} of the individual exposure effects.
\begin{equation}
\label{avg-bias}
\widehat{\text{Bias}}_\tau = \frac{1}{n} \sum_{i=1}^n \left[ \hat{\tau}(\bw_i) - \tau(\bw_i) \right]
\end{equation}
\begin{equation}
\label{rmse}
\widehat{\text{RPEHE}}_\tau = \sqrt{\frac{1}{n} \sum_{i=1}^n \left[ \hat{\tau}(\bw_i) - \tau(\bw_i) \right]^2}
\end{equation}
\begin{equation}
\label{avg-coverage}
\widehat{\text{Coverage}}_\tau = \frac{1}{n} \sum_{i=1}^n \text{I} \left[ \hat{\tau}(\bw_i)_{0.025} \le \tau(\bw_i) \le \hat{\tau}(\bw_i)_{0.975}\right]
\end{equation}
In \eqref{avg-bias}, \eqref{rmse}, and \eqref{avg-coverage}, $\hat{\tau}(\bw_i)$ is the posterior mean individual exposure effect. Results are summarized over 200 simulations for each setting and are presented in Tables \ref{tab:cart-sim-bart-stats} and \ref{tab:friedman-sim-bart-stats}.

\subsection{CART Simulation}

\spacingset{1.1}
\begin{table}[!h]

\caption{\label{tab:cart-sim-bart-stats}CART Simulation Results - BART Predictions}
\centering
\begin{tabular}[t]{cccccc}
\toprule
Type & $M$\textsuperscript{a} & Bias\textsuperscript{b} & RMSE\textsuperscript{b} & Coverage\textsuperscript{b} & Width\textsuperscript{b}\\
\midrule
oracle &  & 0.002 (0.001) & 0.036 (0.001) & 0.940 (0.017) & 0.144 (0.000)\\
\midrule
clbart & 1 & 0.000 (0.001) & 0.067 (0.001) & 0.819 (0.027) & 0.187 (0.003)\\
clbart & 5 & 0.002 (0.001) & 0.056 (0.001) & 0.933 (0.018) & 0.211 (0.002)\\
clbart & 10 & 0.002 (0.001) & 0.058 (0.001) & 0.952 (0.015) & 0.235 (0.002)\\
clbart & 25 & 0.002 (0.001) & 0.063 (0.001) & 0.960 (0.014) & 0.266 (0.001)\\
clbart & 50 & 0.002 (0.001) & 0.069 (0.001) & 0.958 (0.014) & 0.286 (0.001)\\
\bottomrule
\multicolumn{6}{l}{\rule{0pt}{1em}\textsuperscript{a} $M$: Number of trees.}\\
\multicolumn{6}{l}{\rule{0pt}{1em}\textsuperscript{b} Monte Carlo mean and standard errors across 200 simulations reported.}\\
\end{tabular}
\end{table}
\spacingset{1.9}

The oracle shows overall unbiasedness and near 95\% coverage, confirming the validity of the case-crossover design setup (Table \ref{tab:cart-sim-bart-stats}). CL-BART also has negligible bias, but generally has greater RMSE and wider intervals. The latter is to be expected since CL-BART estimates individual (not averaged) effects. RMSE is lowest for the 5 and 10 tree settings, and the average coverage generally increases as the number of trees is increased. Bias and coverage of the confounders is on par with the oracle (see the Supplementary Materials for results). 

The WAIC is lower for the 5, 10, and 25 tree settings for the default tree regularization priors, suggesting the potential for using WAIC to select hyperparameters (see the Supplementary Materials).

\begin{figure}
    \centering
    \includegraphics[width = 0.9\textwidth]{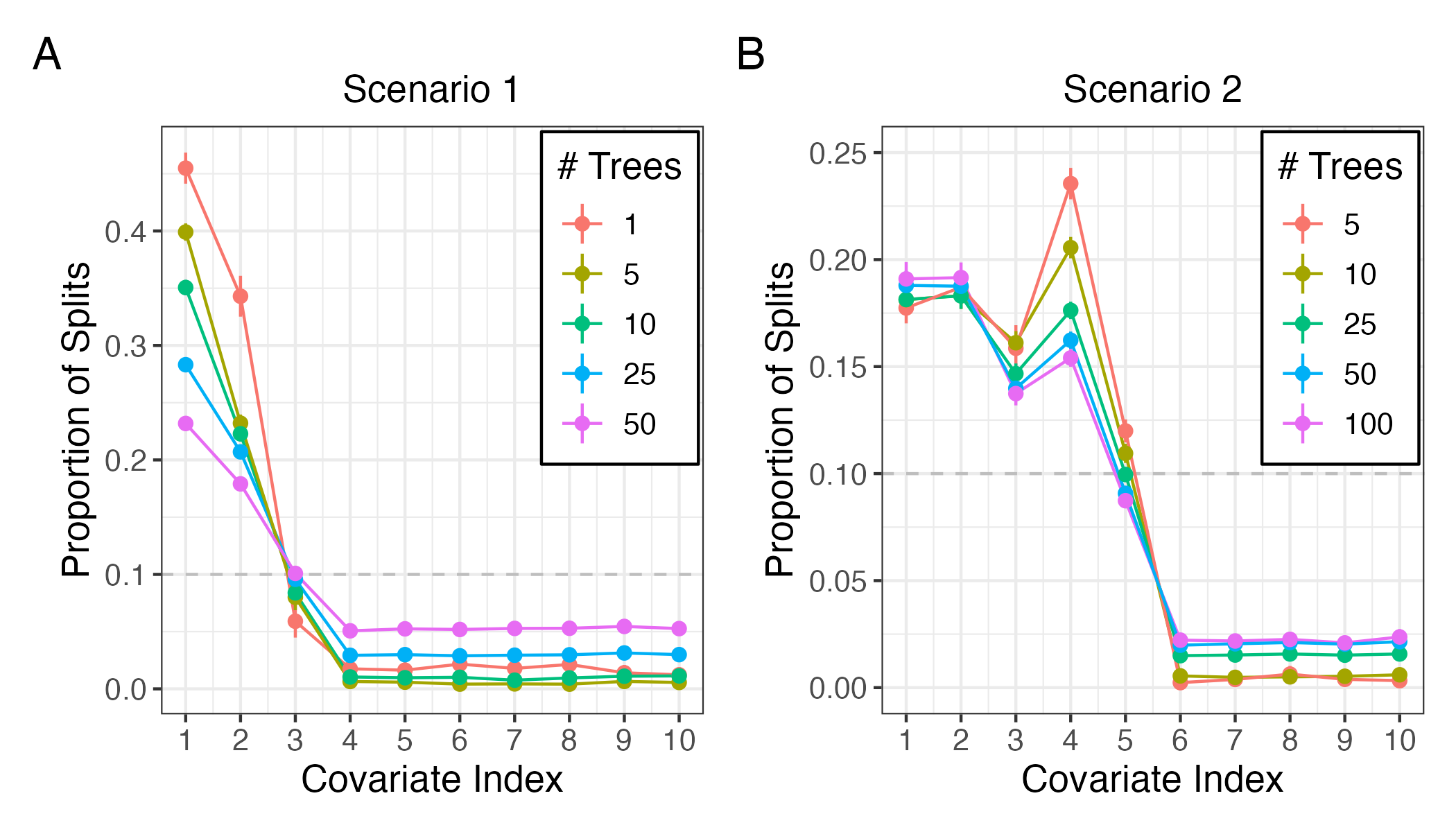}
    \caption{\textbf{Simulation Variable Importance}: Plot of observed split proportions across 200 simulations (Monte Carlo mean and 95\% uncertainty interval presented). Panel A corresponds to Scenario 1 (CART) and Panel B corresponds to Scenario 2 (Friedman).}
    \label{fig:both-sim-var-imp}
\end{figure}

Across all simulations, the important covariates ($w_1', w_2', w_3'$) are typically split on with greater frequency than the remaining seven covariates (Figure \ref{fig:both-sim-var-imp}A). While these values are not perfect indicators of variable importance, this trend suggests the Dirichlet hyperprior is at least somewhat effective at selecting important covariates.

\subsection{Friedman Simulation}
For the Friedman scenario, the oracle achieves low bias and near 95\% average coverage. CL-BART is unbiased even in small ensembles (Table \ref{tab:friedman-sim-bart-stats}). As more trees are added, RMSE and average coverage improve, but interval widths increase. Once again, this is likely due to CL-BART making predictions on the individual level. Estimates of the confounders exhibit low bias and good coverage, and the WAIC for this scenario suggests that larger ensembles perform better, but the improvements diminish as the number of trees approaches 100 (see the Supplementary Materials).

\spacingset{1.1}
\begin{table}[!h]

\caption{\label{tab:friedman-sim-bart-stats}Friedman Simulation Results - BART Predictions}
\centering
\begin{tabular}[t]{cccccc}
\toprule
Type & $M$\textsuperscript{a} & Bias\textsuperscript{b} & RMSE\textsuperscript{b} & Coverage\textsuperscript{b} & Width\textsuperscript{b}\\
\midrule
oracle &  & -0.001 (0.001) & 0.040 (0.001) & 0.949 (0.016) & 0.160 (0.000)\\
\midrule
clbart & 5 & -0.001 (0.001) & 0.165 (0.001) & 0.801 (0.028) & 0.431 (0.002)\\
clbart & 10 & -0.001 (0.001) & 0.144 (0.001) & 0.914 (0.020) & 0.502 (0.002)\\
clbart & 25 & -0.001 (0.001) & 0.130 (0.001) & 0.967 (0.013) & 0.568 (0.002)\\
clbart & 50 & -0.001 (0.001) & 0.127 (0.001) & 0.978 (0.010) & 0.596 (0.003)\\
clbart & 100 & -0.001 (0.001) & 0.126 (0.001) & 0.980 (0.010) & 0.600 (0.003)\\
\bottomrule
\multicolumn{6}{l}{\rule{0pt}{1em}\textsuperscript{a} $M$: Number of trees.}\\
\multicolumn{6}{l}{\rule{0pt}{1em}\textsuperscript{b} Monte Carlo mean and standard errors across 200 simulations reported.}\\
\end{tabular}
\end{table}
\spacingset{1.9}

We see that the important covariates $(w_1, w_2, w_3, w_4, w_5)$ are all split on with greater frequencies, on average, than the remaining covariates (Figure \ref{fig:both-sim-var-imp}B). The Dirichlet hyperprior is particularly effective in this setting since there are many available splitting points for all covariates. Also, CL-BART does well to capture the true marginal partial dependence for each covariate (see the Supplementary Materials). 


\section{Application}
\label{sec:application}

\subsection{Descriptive Statistics}
There were 633,639 ED visits with an AD diagnosis reported during the study period. Patient sex was not reported for 62 cases, race was not reported for 7,662 cases, and ethnicity was not reported for 8,930 cases. Further, only 72,413 cases contained a heat wave per our definition within their referent window (most occurring in the summer months), and thus are the only cases which may be used for estimating heat wave effects. Dropping these cases and implementing the time-stratified case-crossover design resulted in a total of 71,020 cases (319,336 observations).

\spacingset{1.1}
\begin{table}[t]

\caption{\label{tab:app-overall-descriptives}Descriptive Statistics for AD ED Patients, CA 2005-2015}
\centering
\fontsize{12}{14}\selectfont
\begin{tabular}[t]{lc}
\toprule
Characteristic & Overall\\
\midrule
N$^a$ & 71,020\\
\addlinespace[2pt]
Race/Ethnicity$^b$ & \\
\hspace{1em}Hispanic & 11,959 (16.8\%)\\
\hspace{1em}Non-Hispanic White & 46,019 (64.8\%)\\
\hspace{1em}Non-Hispanic Black & 5,635 (7.9\%)\\
\hspace{1em}Non-Hispanic Asian and Pacific Islander & 5,521 (7.8\%)\\
\hspace{1em}Non-Hispanic Other & 1,886 (2.7\%)\\
\addlinespace[2pt]
Sex$^b$ & \\
\hspace{1em}Male & 25,762 (36.3\%)\\
\hspace{1em}Female & 45,258 (63.7\%)\\
\addlinespace[2pt]
Age, yrs$^c$ & 84 (79, 89)\\
\addlinespace[2pt]
Number of Comorbid Conditions$^c$ & 2 (1, 3)\\
\addlinespace[2pt]
Congestive Heart Failure (CHF)$^b$ & 11,494 (16.2\%)\\
\addlinespace[2pt]
Chronic Kidney Disease (CKD)$^b$ & 17,937 (25.3\%)\\
\addlinespace[2pt]
Chronic Obstructive Pulmonary Disease (COPD)$^b$ & 8,483 (11.9\%)\\
\addlinespace[2pt]
Depression (DEP)$^b$ & 9,005 (12.7\%)\\
\addlinespace[2pt]
Diabetes (DIAB)$^b$ & 17,654 (24.9\%)\\
\addlinespace[2pt]
Hypertension (HT)$^b$ & 46,281 (65.2\%)\\
\addlinespace[2pt]
Hyperlipidemia (HLD)$^b$ & 21,575 (30.4\%)\\
\addlinespace[2pt]
\bottomrule
\multicolumn{2}{l}{\rule{0pt}{1em}\textsuperscript{a} N;  \textsuperscript{b} N (\%);  \textsuperscript{c} Median (IQR).}\\
\end{tabular}
\end{table}
\spacingset{1.9}

The sample is primarily non-Hispanic White (64.8\%) and female (63.7\%). The median age is 84 years (IQR: 79, 89). The median number of comorbid conditions is 2 (IQR: 1, 3), and hypertension is the most prevalent condition (65.2\%) (Table \ref{tab:app-overall-descriptives}). Over half of the sample has multiple conditions (56.1\%), with the most common pairings being hypertension and hyperlipidemia (25.0\%), hypertension and CKD (20.7\%), hypertension and diabetes (19.6\%), and hypertension and CHF (12.4\%) .

\subsection{Model Considerations}
Previous studies have found that associations between heat waves and health outcomes may differ by race and ethnicity \citep{madrigano_case-only_2015, knowlton_2006_2009}, so we stratify the analysis into 5 subgroups: Hispanic, non-Hispanic White, non-Hispanic Black, non-Hispanic Asian and Pacific Islander, and non-Hispanic ``other". While we could combine these subgroups and include race as a potential effect moderator, performing the analysis this way effectively forces a split on race first, and ensures fair predictions for each subgroup. A combined analysis would allow subgroups to ``borrow" from each other, but this could potentially mask heterogeneity within smaller subgroups. We do, however, include sex and age as effect moderators alongside the comorbid conditions, with age being the only continuous moderator. The intuition behind including age was to allow it to serve as a proxy for other conditions that are not among those collected. The distribution of sex and age is fairly consistent across the subgroups, but the prevalence of the comorbid conditions may vary (e.g., hypertension and diabetes are less prevalent among non-Hispanic Whites) (See the Supplementary Materials).

On the confounder side, both the daily average temperature and daily average dew-point temperature are modeled using natural cubic splines with four degrees of freedom. Federal holidays are included as a single indicator variable.

We fit a CL-BART model within each subgroup using the following hyperparameter settings: $T = 25$, $k = 1$, $\gamma = 0.95$, and $\xi = 2$. The WAIC was generally similar across different settings, so we only present the results for these particular values. Each chain is run for 10,000 iterations, setting aside the first 5,000 as burn-in and only keeping every fifth sample, resulting in a total of 1,000 posterior samples per subgroup. 

\subsection{Results}
Within each subgroup, there are varying levels of evidence of heterogeneity in the estimated exposure effect. Estimates of the average exposure effect $\bar{\tau}$ are similar to what one would obtain had they ignored effect heterogeneity entirely and simply fit a conditional logistic regression model as specified in \eqref{clr-lik} (Table \ref{tab:clr-versus-clbart}). We also note that the WAIC is similar or better for the CL-BART model in all subgroups (Table \ref{tab:clr-versus-clbart}), suggesting that the overall fit of the models are improved by considering effect heterogeneity, but the additional complexity introduced by using BART may limit generalizability to new data. Density plots of the individual exposure effects illustrate the heterogeneity captured by CL-BART (see the Supplementary Materials). 

\spacingset{1.1}
\begin{table}[H]

\caption{\label{tab:clr-versus-clbart}Homogeneous vs. Average Heterogeneous Estimate for Heat Wave Effect}
\centering
\fontsize{12}{14}\selectfont
\begin{tabular}[t]{lcccc}
\toprule
\multicolumn{1}{c}{ } & \multicolumn{2}{c}{CLR} & \multicolumn{2}{c}{CL-BART} \\
\cmidrule(l{3pt}r{3pt}){2-3} \cmidrule(l{3pt}r{3pt}){4-5}
Subgroup & $\exp{ \left( \hat{\tau} \right)}^a$ (95\% CI) & WAIC & $\exp{ \left( \bar{\tau} \right)}^b$ (95\% CrI) & WAIC\\
\midrule
Hispanic & 0.98 (0.91, 1.06) & 36170 & 0.99 (0.92, 1.05) & 35724\\
Non-Hispanic API & 0.99 (0.89, 1.09) & 16512 & 0.99 (0.90, 1.08) & 16513\\
Non-Hispanic Black & 1.09 (0.97, 1.21) & 16874 & 1.07 (0.96, 1.21) & 16868\\
Non-Hispanic Other & 0.90 (0.73, 1.08) & 5658 & 0.92 (0.76, 1.09) & 5657\\
Non-Hispanic White & 1.03 (0.99, 1.07) & 137904 & 1.01 (0.98, 1.05) & 137894\\
\bottomrule
\multicolumn{5}{l}{\rule{0pt}{1em}API: Asian and Pacific Islander.}\\
\multicolumn{5}{l}{\rule{0pt}{1em}CLR: Conditional Logistic Regression.}\\
\multicolumn{5}{l}{\rule{0pt}{1em}CI: Confidence Interval. CrI: Posterior Credible Interval.}\\
\multicolumn{5}{l}{\rule{0pt}{1em}\textsuperscript{a} Estimated odds-ratio from frequentist CLR with no effect moderators.}\\
\multicolumn{5}{l}{\rule{0pt}{1em}\textsuperscript{b} Average exposure effect from CL-BART model.}\\
\end{tabular}
\end{table}

\spacingset{1.9}

\begin{figure}[H]
    \centering
    \includegraphics[width=0.8\textwidth]{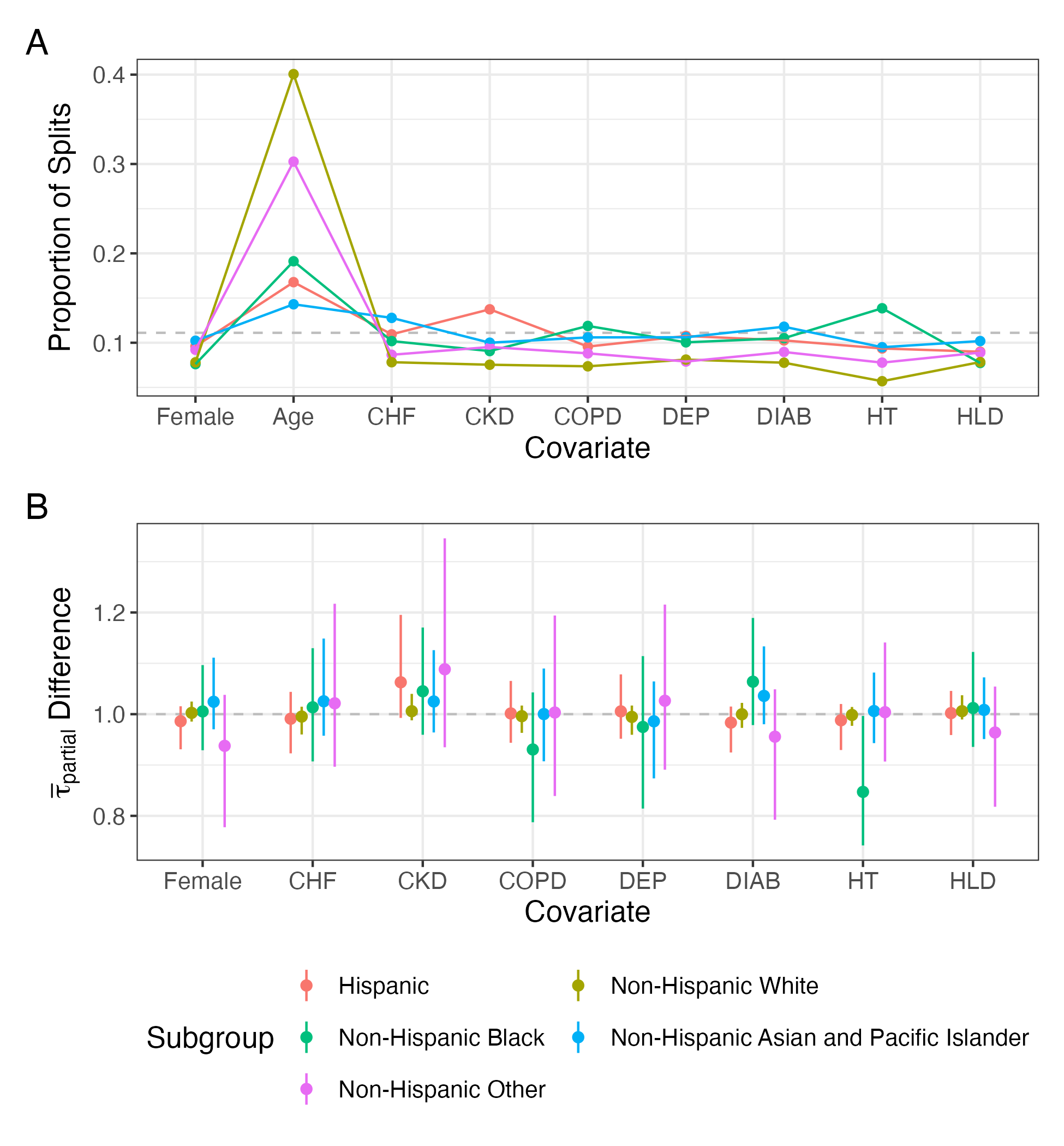}
    \caption{\textbf{Variable Importance and Marginal Partial Dependence}: Panel A displays the proportion of splits in the CL-BART model based on each covariate. Panel B displays the difference in partial average exposure effects for each covariate, except for age (posterior means and 95\% credible intervals presented). Numeric values corresponding to the estimates in panel B are provided in the Supplementary Materials. Abbreviations: DEP: depression, DIAB: diabetes, HT: hypertension, HLD: hyperlipidemia.}
    \label{fig:app-var-imp-and-marg-pd-diff}
\end{figure}

To begin to understand what contributes to this heterogeneity, we visualize the proportions of splits attributable to each moderator in Figure \ref{fig:app-var-imp-and-marg-pd-diff}A. Unsurprisingly, age is split on with greater frequency than any of the binary moderators since it has more available splitting values. However, we also notice that in some cases, certain binary covariates are split on more often than others. Notably, CKD appears to be more important for Hispanics, and hypertension status appears to be more important for Non-Hispanic Blacks. Additionally, we present the marginal contributions (difference in the marginal $\bar{\tau}_{partial}$) for each binary covariate in Figure \ref{fig:app-var-imp-and-marg-pd-diff}B. These estimates are ratios of ORs, and thus represent the multiplicative effect associated with the given moderator on the underlying OR estimate for the association between ED visits and heat waves. For example, the presence of CKD among Hispanics appears to be associated with a harmful modification of the exposure effect for those who have CKD. Similarly, the presence of hypertension among Non-Hispanic Blacks is associated with a protective modification of the exposure effect for those with the condition. While the harmful effect of CKD is most pronounced among the Hispanic subgroup, the estimated OR is positive across all subgroups. Other covariates have mixed effects on the heat wave effect across groups, but these are the most notable.

\begin{figure}[H]
    \centering
    \includegraphics[width = 0.95\textwidth]{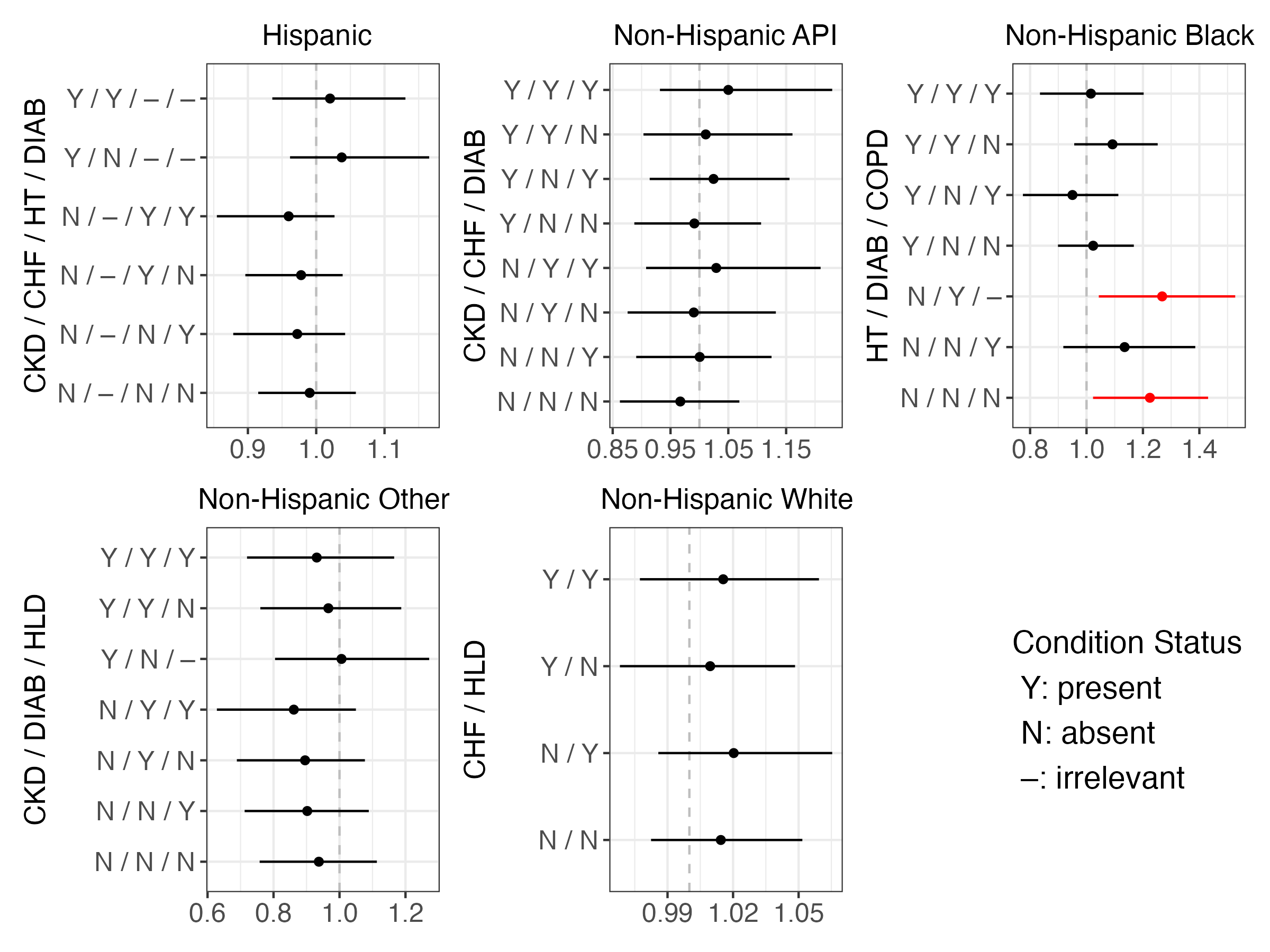}
    \caption{\textbf{CART-Informed Partial Average Exposure Effects}: Posterior mean and 95\% credible intervals for the partial average exposure effects within each leaf of the lower-dimensional CART summaries. Numeric values and CART diagrams are provided in the Supplementary Materials. Abbreviations: DIAB: diabetes, HT: hypertension, HLD: hyperlipidemia.}
    \label{fig:app-cart-pd}
\end{figure}

In order to explore interaction effects estimated via CL-BART, we fit basic CART models using the \texttt{rpart} package \citep{therneau_rpart_2022} to obtain lower-dimensional summaries of the posterior mean individual exposure effects \citep{woody_model_2021} (see the Supplementary Materials). For these models, we drop age and sex from the list of predictors to see how well the heterogeneity can be described by the comorbid conditions alone. We then compute $\bar{\tau}_{partial}$ for each leaf node represented in the summary for each subgroup and plot the results in Figure \ref{fig:app-cart-pd}. These plots are helpful in that they allow one to view the actual exposure effect, as opposed to just ratios of exposure effects in Figure \ref{fig:app-var-imp-and-marg-pd-diff}B.

In Figure \ref{fig:app-cart-pd}, we observe that diabetes appears in 4 out of 5 subgroups, while CKD and CHF are the next most common moderators. Depression is the only condition that did not show up in any of the summaries. The subgroup with the most pronounced risks is Non-Hispanic Black. Among patients in this subgroup, the estimated association between ED visit and heat wave is greatest for those without hypertension, though an additional diagnosis of COPD may reduce this. Findings such as this illustrate the importance of considering interaction among moderators when modeling effect heterogeneity.

Unfortunately, the nice interpretations of the CART summaries come at a cost. The summary $R^2$ \citep{woody_model_2021} for the CART summaries are relatively high for the Hispanic, Non-Hispanic Asian and Pacific Islander, and Non-Hispanic Black subgroups (0.89, 0.71, and 0.79, respectively), but suggest that much of the finer interactions estimated by CL-BART are not captured. The summary $R^2$ is very low for the Non-Hispanic Other (0.39) and Non-Hispanic White (0.09) subgroups. Age was split on with much greater frequency for these subgroups (Figure \ref{fig:app-var-imp-and-marg-pd-diff}A), so in addition to struggling to summarise the CL-BART model fit, it is possible that age and/or sex are the drivers of effect heterogeneity in these subgroups. These findings should not dissuade one from studying effect heterogeneity, but they do illustrate the limitation of automating the analysis.

When fitting the CL-BART model, we monitor trace plots for $\sigma_\mu$, $\bar{\tau}$, and the average number of nodes to ensure adequate mixing and convergence in the final model fits. Examples of these plots are included in the Supplementary Materials

\section{Discussion}
\label{sec:discussion}
CL-BART is a helpful tool for estimating heterogeneous effects in the case-crossover study design commonly used in environmental epidemiology. The primary benefit of CL-BART is its ability to detect and estimate important high-order interactions and functional forms of potential effect moderators without requiring prespecification. Interpreting the heterogeneous effects can be challenging, but the proposed strategies revolving around variable importance, partial dependence, and lower-dimensional summaries provide a good start.

In terms of the application, the most consistent finding across most subgroups was that CKD and/or diabetes may modify the response to heat waves among people with AD. Specifically, having CKD was generally associated with an increased risk of ED visit during heat waves, which aligns with previous work that has established an association between kidney-related illness and extreme heat \citep{johnson_climate_2019, liu_hot_2021, hansen_effect_2008}. Another key finding was the protective-leaning effect of hypertension, particularly among Non-Hispanic Black people with AD. This finding aligns well with previous studies of extreme heat both in California \citep{sherbakov_ambient_2018} and New York \citep{lin_extreme_2009}, and might be attributed to blood vessel dilation in hot weather, decreasing the risk of hypertension-associated morbidity \citep{barnett_effect_2007}. We suspect that a mixture of biological and behavioral changes experienced by and medication(s) taken by those living with these comorbid conditions may be a contributing factor. Finally, while we used co-diagnosis codes at the ED visit to define comorbid conditions, other studies may consider other sources to ascertain pre-existing chronic conditions (e.g., medication use or medical history), to reduce classification error. 

We acknowledge room for future improvements to CL-BART. The computation time is the largest limitation at this point in time. The bottleneck is the repeated evaluation of the conditional logistic regression likelihood required for Fisher scoring when determining the proposal distribution $G$ of interim leaf node parameters and the adaptive rejection sampling of the final leaf node parameters. Exploring additional tree proposals such as those described in \cite{pratola_efficient_2016} and \cite{maia_gp-bart_2022} may improve mixing of the posterior chains, and thus indirectly reduce computation time by lowering the number of MCMC iterations required to reach the stationary distribution. Average runtime for the simulation studies and application analyses are presented in the Supplementary Materials.

Additionally, we have shown that the exposure may either be binary (case study) or continuous (simulations). In both cases, a linear exposure-response relationship is assumed. There are many scenarios where this assumption may be violated. For example, we struggled to achieve model convergence in an exploratory analysis of the ED visit data using continuous daily average temperature as the exposure (results not shown). We suspect a nonlinear exposure-response relationship is at least partly responsible. Extending the CL-BART model to allow for polynomial, splines, and other flexible functions of the exposure could be desirable. Each basis function would require its own forest to be managed, substantially increasing the computational burden of an already burdensome algorithm.

In conclusion, CL-BART serves as a robust alternative to the typical strategies for estimating heterogeneous effects in the case-crossover design, using RJMCMC to integrate the flexibility of BART with traditional conditional logistic regression. This framework
offers researchers a powerful tool to disentangle and model heterogeneous effects, whether it be in the context of treatment outcomes, environmental exposures, or any other one-to-many matched case-control study.








\newpage

\bibliographystyle{apalike}
\bibliography{references}

\end{document}


\def\spacingset#1{\renewcommand{\baselinestretch}%
{#1}\small\normalsize} \spacingset{1}


\if1\blind
{
  \title{\bf Supplementary Material for \\Estimating Heterogeneous Exposure Effects in the Case-Crossover Design using BART}
  \author{Jacob R. Englert\\
    Department of Biostatistics and Bioinformatics, Emory University\\[1ex]
    Stefanie T. Ebelt \\
    Department of Environmental Health, Emory University \\[1ex]
    Howard H. Chang \\
    Department of Biostatistics and Bioinformatics, Emory University}
  \maketitle
} \fi

\if0\blind
{
  \bigskip
  \bigskip
  \bigskip
  \begin{center}
    {\LARGE\bf Supplementary Material for \\Estimating Heterogeneous Exposure Effects in the Case-Crossover Design using BART}
\end{center}
  \medskip
} \fi

\spacingset{1.9} 

\appendix
\renewcommand\thefigure{S.\arabic{figure}}
\renewcommand\thetable{S.\arabic{table}}
\renewcommand\thesection{S.\arabic{section}}
\setcounter{table}{0}
\setcounter{figure}{0}

\section{List of Comorbid Condition Diagnosis Codes}
We used both primary and secondary International Classification of Diseases (ICD) diagnosis codes to identify AD ED visits and comorbid conditions including chronic kidney disease, chronic obstructive pulmonary disease, congestive heart failure, depression, diabetes, hypertension, and hyperlipidemia. The exact codes are provided in Table \ref{tab:icd-codes}. ICD-9 codes are used for records observed prior to October 1, 2015.

\spacingset{1}
\begin{landscape}\begingroup\fontsize{10}{12}\selectfont

\begin{longtable}[t]{p{3cm} p{7cm} p{9cm}}
\caption{\label{tab:icd-codes}International Classification of Diseases (ICD) codes used to identify AD ED visits and other comorbid conditions.}\\
\toprule
Condition & ICD-9 Codes & ICD-10 Codes\\
\midrule
\endfirsthead
\caption[]{International Classification of Diseases (ICD) codes used to identify AD ED visits and other comorbid conditions. \textit{(continued)}}\\
\toprule
Condition & ICD-9 Codes & ICD-10 Codes\\
\midrule
\endhead
\midrule
\multicolumn{3}{r@{}}{\textit{(continued \ldots)}}\
\endfoot
\bottomrule
\endlastfoot
Alzheimer's Disease & 331.0 & G30.0, G30.1, G30.8, G30.9\\
\addlinespace
Chronic Kidney Disease & 016.00, 016.01, 016.02, 016.03, 016.04, 016.05, 016.06, 095.4, 189.0, 189.9, 223.0, 236.91, 249.40, 249.41, 250.40, 250.41, 250.42, 250.43, 271.4, 274.10, 283.11, 403.01, 403.11, 403.91, 404.02, 404.03, 404.12, 404.13, 404.92, 404.93, 440.1, 442.1, 572.4, 580.0, 580.4, 580.81, 580.89, 580.9, 581.0, 581.1, 581.2, 581.3, 581.81, 581.89, 581.9, 582.0, 582.1, 582.2, 582.4, 582.81, 582.89, 582.9, 583.0, 583.1, 583.2, 583.4, 583.6, 583.7, 583.81, 583.89, 583.9, 584.5, 584.6, 584.7, 584.8, 584.9, 585.1, 585.2, 585.3, 585.4, 585.5, 585.6, 585.9, 586, 587, 588.0, 588.1, 588.81, 588.89, 588.9, 591, 753.12, 753.13, 753.14, 753.15, 753.16, 753.17, 753.19, 753.20, 753.21, 753.22, 753.23, 753.29, 794.4 & A18.11, A52.75, B52.0, C64.1, C64.2, C64.9, C68.9, D30.00, D30.01, D30.02, D41.00, D41.01, D41.02, D41.10, D41.11, D41.12, D41.20, D41.21, D41.22, D59.3, E08.21, E08.22, E08.29, E08.65, E09.21, E09.22, E09.29, E10.21, E10.22, E10.29, E10.65, E11.21, E11.22, E11.29, E11.65, E13.21, E13.22, E13.29, E74.8, I12.0, I12.9, I13.0, I13.10, I13.11, I13.2, I70.1, I72.2, K76.7, M10.30, M10.311, M10.312, M10.319, M10.321, M10.322, M10.329, M10.331, M10.332, M10.339, M10.341, M10.342, M10.349, M10.351, M10.352, M10.359, M10.361, M10.362, M10.369, M10.371, M10.372, M10.379, M10.38, M10.39, M32.14, M32.15, M35.04, N00.0, N00.1, N00.2, N00.3, N00.4, N00.5, N00.6, N00.7, N00.8, N00.9, N00.A, N01.0, N01.1, N01.2, N01.3, N01.4, N01.5, N01.6, N01.7, N01.8, N01.9, N01.A, N02.0, N02.1, N02.2, N02.3, N02.4, N02.5, N02.6, N02.7, N02.8, N02.9, N02.A, N03.0, N03.1, N03.2, N03.3, N03.4, N03.5, N03.6, N03.7, N03.8, N03.9, N03.A, N04.0, N04.1, N04.2, N04.3, N04.4, N04.5, N04.6, N04.7, N04.8, N04.9, N04.A, N05.0, N05.1, N05.2, N05.3, N05.4, N05.5, N05.6, N05.7, N05.8, N05.9, N05.A, N06.0, N06.1, N06.2, N06.3, N06.4, N06.5, N06.6, N06.7, N06.8, N06.9, N06.A, N07.0, N07.1, N07.2, N07.3, N07.4, N07.5, N07.6, N07.7, N07.8, N07.9, N07.A, N08, N13.1, N13.2, N13.30, N13.39, N14.0, N14.1, N14.2, N14.3, N14.4, N15.0, N15.8, N15.9, N16, N17.0, N17.1, N17.2, N17.8, N17.9, N18.1, N18.2, N18.3, N18.30, N18.31, N18.32, N18.4, N18.5, N18.6, N18.9, N19, N25.0, N25.1, N25.81, N25.89, N25.9, N26.1, N26.9, Q61.02, Q61.11, Q61.19, Q61.2, Q61.3, Q61.4, Q61.5, Q61.8, Q62.0, Q62.2, Q62.10, Q62.11, Q62.12, Q62.31, Q62.32, Q62.39, R94.4\\
\addlinespace
Chronic Obstructive Pulmonary Disease & 491, 492, 496 & J41, J42, J43, J44\\
\addlinespace
Congestive Heart Failure & 428 & I42, I50, I51\\
\addlinespace
Depression & 296.20, 296.21, 296.22, 296.23, 296.24, 296.25, 296.26, 296.30, 296.31, 296.32, 296.33, 296.34, 296.35, 296.36, 296.51, 296.52, 296.53, 296.54, 296.55, 296.56, 296.60, 296.61, 296.62, 296.63, 296.64, 296.65, 296.66, 296.89, 298.0, 300.4, 309.1, 311 & F31.30, F31.31, F31.32, F31.4, F31.5, F31.60, F31.61, F31.62, F31.63, F31.64, F31.75, F31.76, F31.77, F31.78, F31.81, F32.0, F32.1, F32.2, F32.3, F32.4, F32.5, F32.9, F33.0, F33.1, F33.2, F33.3, F33.40, F33.41, F33.42, F33.8, F33.9, F34.1, F43.21, F43.23\\
\addlinespace
Diabetes & 249, 250 & E08, E09, E10, E11, E12, E13\\
\addlinespace
Hyperlipidemia & 272.0, 272.1, 272.2, 272.3, 272.4 & E78.0, E78.00, E78.01, E78.1, E78.2, E78.3, E78.4, E78.41, E78.49, E78.5\\
\addlinespace
Hypertension & 401, 402, 403, 404, 405 & I10, I11, I12, I13, I15\\*
\end{longtable}
\endgroup{}
\end{landscape}
\spacingset{1.9}

\newpage

\section[The Proposal Distribution]{The Proposal Distribution $\displaystyle G$}
\label{supp:prop-dist}

In the reversible jump algorithm, the choice of the proposal distributions $G_{grow}$, $G_{prune}$, and $G_{change}$ is crucial to ensuring good mixing. \cite{linero_generalized_2024} suggest a Normal proposal distribution based on the Laplace approximation. For any node $\eta$ that is involved in a tree update proposal, we may sample $\mu_\eta \sim \text{Normal}(m_\eta,v_\eta^2)$, where

$$m_\eta = \text{arg max}_\mu \sum_{i:\bw_i \mapsto \eta} \log \pi(\by_i \mid \lambda_i^r + \mu, \bbeta) + \log \pi_\mu(\mu)$$
$$v_\eta^{-2} = \sum_{i:\bw_i \mapsto \eta} \cI(\lambda_i^r + m_\eta, \bbeta) - \frac{d^2}{d\mu^2}\log \pi_\mu(\mu) \mid_{\mu = m_\eta}$$

Both $m_\eta$ and $v_\eta^{-2}$ may obtained using a Fisher scoring algorithm. Details for the gradient and Fisher information computation required in this step are provided below in Section \ref{supp:clr-deriv}. For a starting point, we use the parameter from the leaf node that was split in either a grow or change move, or a weighted average of the leaf node parameters that were deleted in a prune move.

We have found that this strategy typically works well, but in certain cases these starting values may be inadequate, causing the Fisher scoring algorithm to not converge. This usually occurs when there exists an extreme imbalance in the covariate space or the true predictions for leaf nodes are very different from that of their parent. In this rare scenario, we estimate $m_\eta$ by first maximizing the likelihood using frequentist conditional logistic regression, followed by an application of the optimization technique introduced in \citet{brent_algorithms_1972} to incorporate prior information during the maximization.

\subsection{Conditional Logistic Regression Derivations}
\label{supp:clr-deriv}

The conditional likelihood of a case-crossover design for $i = 1, \ldots, n$ individuals with one time-varying exposure of interest $z_{it}$, several time-varying confounding variables $\bx_{it}$, several constant-within-individual characteristics $\bw_i$ that modify the exposure-response relationship, referent windows $\cW_i$, using the fact that $\sum_{t \in \cW_i}Y_{it} = 1$, is given by:

\begin{equation}
\pi(\by \mid \tau(\bw), \bbeta) = \prod_{i=1}^n \pi(\by_i \mid \tau(\bw_i), \bbeta) = \prod_{i=1}^n \frac{\exp \left\{\bx_{it_i}\bbeta + \tau(\bw_i) z_{it_i} \right\}}{\sum_{t \in \cW_i} \exp \left\{\bx_{it}\bbeta + \tau(\bw_i) z_{it} \right\}}    
\end{equation}

The log-likelihood is given by:

\begin{equation}
l(\by \mid \tau(\bw), \bbeta) = \sum_{i=1}^n \left[ \bx_{it_i}\bbeta + \tau(\bw_i) z_{it_i} - \log \sum_{t \in \cW_i} \exp \left\{\bx_{it}\bbeta + \tau(\bw_i) z_{it} \right\} \right]    
\end{equation}

The score with respect to $\tau(\bw_i)$ is given by:

\begin{equation}
U(\by \mid \tau(\bw)) = \sum_{i=1}^n \left[ z_{it_i} - \frac{\sum_{t \in \cW_i} z_{it} \exp \left\{\bx_{it}\bbeta + \tau(\bw_i) z_{it} \right\}}{\sum_{t \in \cW_i} \exp \left\{\bx_{it}\bbeta + \tau(\bw_i) z_{it} \right\}} \right]    
\end{equation}

Letting $s_{it} = \exp \left\{ \bx_{it}\bbeta + \tau(\bw_i) z_{it} \right\}$, the Fisher information is given by:

\begin{equation}
I(\by \mid \tau(\bw)) = \sum_{i=1}^n \left[ \frac{\left( \sum_{t \in \cW_i} s_{it} \right) \left( \sum_{t \in \cW_i} z_{it}^2 s_{it} \right) - \left( \sum_{t \in \cW_i} z_{it} s_{it} \right)^2}{\left( \sum_{t \in \cW_i} s_{it} \right)^2} \right]
\end{equation}

Note that in this scenario the Fisher information is equal to the observed information.

\newpage

\section{Additional Simulation Results}
\label{supp:sim}

\spacingset{1.1}
\begin{table}[!h]

\caption{\label{tab:ext-cart-sim-bart-stats}Extended CART Simulation Results - BART Predictions}
\centering
\fontsize{11}{13}\selectfont
\begin{tabular}[t]{cccccccc}
\toprule
Type & $M$\textsuperscript{a} & $k$\textsuperscript{b} & $(\gamma, \xi)$\textsuperscript{c} & Bias\textsuperscript{d} & RMSE\textsuperscript{d} & Coverage\textsuperscript{d} & Width\textsuperscript{d}\\
\midrule
oracle &  &  &  & 0.002 (0.001) & 0.036 (0.001) & 0.940 (0.017) & 0.144 (0.000)\\
\midrule
\addlinespace[2pt]
clbart & 1 & 0.1 & (0.5, 3) & -0.001 (0.001) & 0.070 (0.002) & 0.766 (0.030) & 0.170 (0.002)\\
clbart & 1 & 0.1 & (0.95, 2) & -0.001 (0.001) & 0.069 (0.001) & 0.813 (0.028) & 0.189 (0.003)\\
clbart & 1 & 0.5 & (0.5, 3) & 0.000 (0.001) & 0.069 (0.002) & 0.771 (0.030) & 0.168 (0.002)\\
clbart & 1 & 0.5 & (0.95, 2) & 0.000 (0.001) & 0.069 (0.001) & 0.807 (0.028) & 0.189 (0.003)\\
clbart & 1 & 1.0 & (0.5, 3) & 0.000 (0.001) & 0.068 (0.001) & 0.762 (0.030) & 0.164 (0.002)\\
clbart & 1 & 1.0 & (0.95, 2) & 0.000 (0.001) & 0.067 (0.001) & 0.819 (0.027) & 0.187 (0.003)\\
\midrule
\addlinespace[2pt]
clbart & 5 & 0.1 & (0.5, 3) & 0.002 (0.001) & 0.058 (0.001) & 0.862 (0.024) & 0.174 (0.002)\\
clbart & 5 & 0.1 & (0.95, 2) & 0.001 (0.001) & 0.055 (0.001) & 0.938 (0.017) & 0.214 (0.002)\\
clbart & 5 & 0.5 & (0.5, 3) & 0.002 (0.001) & 0.059 (0.001) & 0.856 (0.025) & 0.174 (0.002)\\
clbart & 5 & 0.5 & (0.95, 2) & 0.001 (0.001) & 0.056 (0.001) & 0.937 (0.017) & 0.213 (0.002)\\
clbart & 5 & 1.0 & (0.5, 3) & 0.002 (0.001) & 0.059 (0.001) & 0.856 (0.025) & 0.173 (0.002)\\
clbart & 5 & 1.0 & (0.95, 2) & 0.002 (0.001) & 0.056 (0.001) & 0.933 (0.018) & 0.211 (0.002)\\
\midrule
\addlinespace[2pt]
clbart & 10 & 0.1 & (0.5, 3) & 0.002 (0.001) & 0.059 (0.001) & 0.882 (0.023) & 0.185 (0.002)\\
clbart & 10 & 0.1 & (0.95, 2) & 0.001 (0.001) & 0.058 (0.001) & 0.950 (0.015) & 0.235 (0.002)\\
clbart & 10 & 0.5 & (0.5, 3) & 0.002 (0.001) & 0.059 (0.001) & 0.883 (0.023) & 0.184 (0.002)\\
clbart & 10 & 0.5 & (0.95, 2) & 0.002 (0.001) & 0.058 (0.001) & 0.952 (0.015) & 0.236 (0.002)\\
clbart & 10 & 1.0 & (0.5, 3) & 0.002 (0.001) & 0.059 (0.001) & 0.879 (0.023) & 0.183 (0.002)\\
clbart & 10 & 1.0 & (0.95, 2) & 0.002 (0.001) & 0.058 (0.001) & 0.952 (0.015) & 0.235 (0.002)\\
\midrule
\addlinespace[2pt]
clbart & 25 & 0.1 & (0.5, 3) & 0.002 (0.001) & 0.060 (0.001) & 0.900 (0.021) & 0.199 (0.002)\\
clbart & 25 & 0.1 & (0.95, 2) & 0.002 (0.001) & 0.063 (0.001) & 0.955 (0.015) & 0.261 (0.001)\\
clbart & 25 & 0.5 & (0.5, 3) & 0.002 (0.001) & 0.059 (0.001) & 0.902 (0.021) & 0.200 (0.002)\\
clbart & 25 & 0.5 & (0.95, 2) & 0.002 (0.001) & 0.064 (0.001) & 0.959 (0.014) & 0.266 (0.001)\\
clbart & 25 & 1.0 & (0.5, 3) & 0.002 (0.001) & 0.059 (0.001) & 0.907 (0.021) & 0.199 (0.002)\\
clbart & 25 & 1.0 & (0.95, 2) & 0.002 (0.001) & 0.063 (0.001) & 0.960 (0.014) & 0.266 (0.001)\\
\midrule
\addlinespace[2pt]
clbart & 50 & 0.1 & (0.5, 3) & 0.002 (0.001) & 0.061 (0.001) & 0.908 (0.020) & 0.208 (0.001)\\
clbart & 50 & 0.1 & (0.95, 2) & 0.002 (0.001) & 0.069 (0.001) & 0.953 (0.015) & 0.279 (0.001)\\
clbart & 50 & 0.5 & (0.5, 3) & 0.002 (0.001) & 0.061 (0.001) & 0.912 (0.020) & 0.209 (0.001)\\
clbart & 50 & 0.5 & (0.95, 2) & 0.002 (0.001) & 0.069 (0.001) & 0.957 (0.014) & 0.284 (0.001)\\
clbart & 50 & 1.0 & (0.5, 3) & 0.002 (0.001) & 0.060 (0.001) & 0.916 (0.020) & 0.210 (0.001)\\
clbart & 50 & 1.0 & (0.95, 2) & 0.002 (0.001) & 0.069 (0.001) & 0.958 (0.014) & 0.286 (0.001)\\
\bottomrule
\multicolumn{8}{l}{\rule{0pt}{1em}\textsuperscript{a} $M$: Number of trees.}\\
\multicolumn{8}{l}{\rule{0pt}{1em}\textsuperscript{b} $k$: Numerator of scale parameter for half-Cauchy hyper-prior.}\\
\multicolumn{8}{l}{\rule{0pt}{1em}\textsuperscript{c} $(\gamma, \xi)$: Hyperparameters for tree depth prior.}\\
\multicolumn{8}{l}{\rule{0pt}{1em}\textsuperscript{d} Monte Carlo mean and standard errors across 200 simulations reported.}\\
\end{tabular}
\end{table}
\spacingset{1.9}

\newpage
\spacingset{1.1}
\begin{table}[!ht]

\caption{\label{tab:ext-cart-sim-beta-bias}Extended CART Simulation Results - Confounder Estimates (Bias)}
\centering
\fontsize{11}{13}\selectfont
\begin{tabular}[t]{ccccccccc}
\toprule
\multicolumn{4}{c}{ } & \multicolumn{5}{c}{Bias $\times$ 1,000\textsuperscript{d}} \\
\cmidrule(l{3pt}r{3pt}){5-9}
Type & $M$\textsuperscript{a} & $k$\textsuperscript{b} & $(\gamma, \xi)$\textsuperscript{c} & $\beta_1$ & $\beta_2$ & $\beta_3$ & $\beta_4$ & $\beta_5$\\
\midrule
oracle &  &  &  & 6.54 (4.49) & -11.05 (4.77) & 1.59 (5.10) & -3.90 (4.65) & -0.46 (4.69)\\
\midrule
\addlinespace[2pt]
clbart & 1 & 0.1 & (0.5, 3) & 5.60 (4.51) & -11.30 (4.78) & 1.56 (5.13) & -3.85 (4.66) & -0.06 (4.65)\\
clbart & 1 & 0.1 & (0.95, 2) & 5.68 (4.49) & -11.56 (4.80) & 2.08 (5.10) & -4.17 (4.66) & 0.16 (4.66)\\
clbart & 1 & 0.5 & (0.5, 3) & 5.96 (4.50) & -11.90 (4.79) & 1.55 (5.09) & -4.39 (4.67) & 0.31 (4.70)\\
clbart & 1 & 0.5 & (0.95, 2) & 5.53 (4.48) & -11.76 (4.75) & 1.56 (5.10) & -3.80 (4.66) & 0.22 (4.69)\\
clbart & 1 & 1.0 & (0.5, 3) & 6.56 (4.48) & -11.50 (4.77) & 1.46 (5.11) & -4.22 (4.65) & -0.06 (4.64)\\
clbart & 1 & 1.0 & (0.95, 2) & 6.06 (4.51) & -11.43 (4.80) & 1.30 (5.09) & -3.48 (4.66) & 0.07 (4.63)\\
\midrule
\addlinespace[2pt]
clbart & 5 & 0.1 & (0.5, 3) & 5.63 (4.49) & -10.93 (4.79) & 1.64 (5.10) & -4.34 (4.65) & -0.38 (4.71)\\
clbart & 5 & 0.1 & (0.95, 2) & 5.90 (4.50) & -11.43 (4.76) & 1.77 (5.11) & -4.02 (4.65) & 0.20 (4.69)\\
clbart & 5 & 0.5 & (0.5, 3) & 6.17 (4.48) & -11.10 (4.81) & 1.59 (5.11) & -3.98 (4.65) & 0.02 (4.70)\\
clbart & 5 & 0.5 & (0.95, 2) & 5.43 (4.50) & -11.48 (4.76) & 1.42 (5.15) & -3.74 (4.65) & 0.76 (4.69)\\
clbart & 5 & 1.0 & (0.5, 3) & 6.24 (4.50) & -11.31 (4.77) & 1.55 (5.10) & -3.71 (4.67) & 0.30 (4.69)\\
clbart & 5 & 1.0 & (0.95, 2) & 5.86 (4.50) & -11.22 (4.79) & 1.81 (5.11) & -3.64 (4.67) & 0.02 (4.68)\\
\midrule
\addlinespace[2pt]
clbart & 10 & 0.1 & (0.5, 3) & 6.11 (4.46) & -11.37 (4.78) & 0.80 (5.11) & -4.06 (4.69) & 0.25 (4.67)\\
clbart & 10 & 0.1 & (0.95, 2) & 5.95 (4.51) & -11.78 (4.80) & 1.62 (5.13) & -4.00 (4.63) & 0.69 (4.71)\\
clbart & 10 & 0.5 & (0.5, 3) & 6.28 (4.49) & -11.31 (4.80) & 1.78 (5.11) & -4.04 (4.65) & 0.53 (4.71)\\
clbart & 10 & 0.5 & (0.95, 2) & 5.44 (4.49) & -11.57 (4.79) & 0.75 (5.13) & -4.04 (4.66) & 0.74 (4.71)\\
clbart & 10 & 1.0 & (0.5, 3) & 6.16 (4.49) & -11.39 (4.79) & 1.35 (5.14) & -4.21 (4.64) & 0.21 (4.67)\\
clbart & 10 & 1.0 & (0.95, 2) & 5.54 (4.49) & -11.01 (4.82) & 1.86 (5.08) & -3.93 (4.69) & 0.96 (4.69)\\
\midrule
\addlinespace[2pt]
clbart & 25 & 0.1 & (0.5, 3) & 5.78 (4.51) & -11.56 (4.79) & 1.64 (5.11) & -3.41 (4.64) & 0.21 (4.71)\\
clbart & 25 & 0.1 & (0.95, 2) & 5.80 (4.50) & -10.96 (4.81) & 1.66 (5.13) & -3.94 (4.68) & 0.84 (4.72)\\
clbart & 25 & 0.5 & (0.5, 3) & 6.25 (4.51) & -11.64 (4.78) & 1.59 (5.13) & -3.80 (4.68) & 0.23 (4.71)\\
clbart & 25 & 0.5 & (0.95, 2) & 5.19 (4.53) & -11.45 (4.78) & 1.54 (5.10) & -3.87 (4.66) & 0.44 (4.73)\\
clbart & 25 & 1.0 & (0.5, 3) & 5.71 (4.49) & -11.49 (4.80) & 1.72 (5.14) & -3.69 (4.67) & 0.34 (4.70)\\
clbart & 25 & 1.0 & (0.95, 2) & 5.65 (4.51) & -11.34 (4.79) & 1.97 (5.13) & -3.98 (4.63) & 0.86 (4.72)\\
\midrule
\addlinespace[2pt]
clbart & 50 & 0.1 & (0.5, 3) & 5.58 (4.50) & -11.22 (4.80) & 1.60 (5.14) & -3.79 (4.66) & 0.21 (4.71)\\
clbart & 50 & 0.1 & (0.95, 2) & 5.64 (4.56) & -11.51 (4.78) & 1.70 (5.11) & -3.64 (4.68) & 1.06 (4.68)\\
clbart & 50 & 0.5 & (0.5, 3) & 5.86 (4.49) & -11.21 (4.80) & 1.01 (5.09) & -3.96 (4.68) & -0.27 (4.74)\\
clbart & 50 & 0.5 & (0.95, 2) & 5.97 (4.53) & -11.80 (4.77) & 1.47 (5.11) & -3.61 (4.68) & 1.16 (4.68)\\
clbart & 50 & 1.0 & (0.5, 3) & 5.87 (4.55) & -11.56 (4.82) & 1.30 (5.11) & -3.64 (4.64) & 0.40 (4.69)\\
clbart & 50 & 1.0 & (0.95, 2) & 5.33 (4.50) & -11.11 (4.78) & 1.50 (5.13) & -3.90 (4.67) & 0.21 (4.74)\\
\bottomrule
\multicolumn{9}{l}{\rule{0pt}{1em}\textsuperscript{a} $M$: Number of trees.}\\
\multicolumn{9}{l}{\rule{0pt}{1em}\textsuperscript{b} $k$: Numerator of scale parameter for half-Cauchy hyper-prior.}\\
\multicolumn{9}{l}{\rule{0pt}{1em}\textsuperscript{c} $(\gamma, \xi)$: Hyperparameters for tree depth prior.}\\
\multicolumn{9}{l}{\rule{0pt}{1em}\textsuperscript{d} Monte Carlo mean and standard errors across 200 simulations reported.}\\
\end{tabular}
\end{table}
\spacingset{1.9}

\newpage
\spacingset{1.1}
\begin{table}[!ht]

\caption{\label{tab:ext-cart-sim-beta-coverage}Extended CART Simulation Results - Confounder Estimates (Coverage)}
\centering
\fontsize{11}{13}\selectfont
\begin{tabular}[t]{ccccccccc}
\toprule
\multicolumn{4}{c}{ } & \multicolumn{5}{c}{Coverage\textsuperscript{d}} \\
\cmidrule(l{3pt}r{3pt}){5-9}
Type & $M$\textsuperscript{a} & $k$\textsuperscript{b} & $(\gamma, \xi)$\textsuperscript{c} & $\beta_1$ & $\beta_2$ & $\beta_3$ & $\beta_4$ & $\beta_5$\\
\midrule
oracle &  &  &  & 0.95 & 0.93 & 0.92 & 0.95 & 0.95\\
\midrule
\addlinespace[2pt]
clbart & 1 & 0.1 & (0.5, 3) & 0.94 & 0.93 & 0.92 & 0.95 & 0.94\\
clbart & 1 & 0.1 & (0.95, 2) & 0.96 & 0.93 & 0.92 & 0.93 & 0.95\\
clbart & 1 & 0.5 & (0.5, 3) & 0.94 & 0.92 & 0.91 & 0.94 & 0.93\\
clbart & 1 & 0.5 & (0.95, 2) & 0.95 & 0.94 & 0.92 & 0.94 & 0.95\\
clbart & 1 & 1.0 & (0.5, 3) & 0.96 & 0.93 & 0.91 & 0.94 & 0.94\\
clbart & 1 & 1.0 & (0.95, 2) & 0.93 & 0.92 & 0.92 & 0.94 & 0.94\\
\midrule
\addlinespace[2pt]
clbart & 5 & 0.1 & (0.5, 3) & 0.95 & 0.93 & 0.92 & 0.94 & 0.94\\
clbart & 5 & 0.1 & (0.95, 2) & 0.95 & 0.94 & 0.92 & 0.95 & 0.96\\
clbart & 5 & 0.5 & (0.5, 3) & 0.95 & 0.92 & 0.92 & 0.95 & 0.94\\
clbart & 5 & 0.5 & (0.95, 2) & 0.94 & 0.93 & 0.92 & 0.94 & 0.94\\
clbart & 5 & 1.0 & (0.5, 3) & 0.94 & 0.94 & 0.92 & 0.95 & 0.92\\
clbart & 5 & 1.0 & (0.95, 2) & 0.94 & 0.94 & 0.92 & 0.94 & 0.93\\
\midrule
\addlinespace[2pt]
clbart & 10 & 0.1 & (0.5, 3) & 0.95 & 0.92 & 0.90 & 0.94 & 0.95\\
clbart & 10 & 0.1 & (0.95, 2) & 0.95 & 0.92 & 0.90 & 0.95 & 0.94\\
clbart & 10 & 0.5 & (0.5, 3) & 0.93 & 0.93 & 0.92 & 0.93 & 0.95\\
clbart & 10 & 0.5 & (0.95, 2) & 0.94 & 0.93 & 0.92 & 0.94 & 0.94\\
clbart & 10 & 1.0 & (0.5, 3) & 0.95 & 0.92 & 0.91 & 0.94 & 0.95\\
clbart & 10 & 1.0 & (0.95, 2) & 0.95 & 0.93 & 0.91 & 0.94 & 0.95\\
\midrule
\addlinespace[2pt]
clbart & 25 & 0.1 & (0.5, 3) & 0.95 & 0.94 & 0.92 & 0.94 & 0.95\\
clbart & 25 & 0.1 & (0.95, 2) & 0.95 & 0.94 & 0.91 & 0.93 & 0.94\\
clbart & 25 & 0.5 & (0.5, 3) & 0.94 & 0.93 & 0.92 & 0.93 & 0.95\\
clbart & 25 & 0.5 & (0.95, 2) & 0.95 & 0.92 & 0.92 & 0.94 & 0.95\\
clbart & 25 & 1.0 & (0.5, 3) & 0.95 & 0.94 & 0.92 & 0.95 & 0.94\\
clbart & 25 & 1.0 & (0.95, 2) & 0.94 & 0.93 & 0.90 & 0.94 & 0.94\\
\midrule
\addlinespace[2pt]
clbart & 50 & 0.1 & (0.5, 3) & 0.94 & 0.92 & 0.90 & 0.94 & 0.95\\
clbart & 50 & 0.1 & (0.95, 2) & 0.94 & 0.93 & 0.92 & 0.94 & 0.96\\
clbart & 50 & 0.5 & (0.5, 3) & 0.95 & 0.93 & 0.90 & 0.94 & 0.94\\
clbart & 50 & 0.5 & (0.95, 2) & 0.95 & 0.93 & 0.91 & 0.93 & 0.94\\
clbart & 50 & 1.0 & (0.5, 3) & 0.93 & 0.94 & 0.92 & 0.95 & 0.94\\
clbart & 50 & 1.0 & (0.95, 2) & 0.95 & 0.93 & 0.92 & 0.94 & 0.93\\
\bottomrule
\multicolumn{9}{l}{\rule{0pt}{1em}\textsuperscript{a} $M$: Number of trees.}\\
\multicolumn{9}{l}{\rule{0pt}{1em}\textsuperscript{b} $k$: Numerator of scale parameter for half-Cauchy hyper-prior.}\\
\multicolumn{9}{l}{\rule{0pt}{1em}\textsuperscript{c} $(\gamma, \xi)$: Hyperparameters for tree depth prior.}\\
\multicolumn{9}{l}{\rule{0pt}{1em}\textsuperscript{d} 95\% credible interval coverage across 200 simulations.}\\
\end{tabular}
\end{table}
\spacingset{1.9}

\newpage
\begin{figure}[H]
    \centering
    \includegraphics[width=0.95\textwidth]{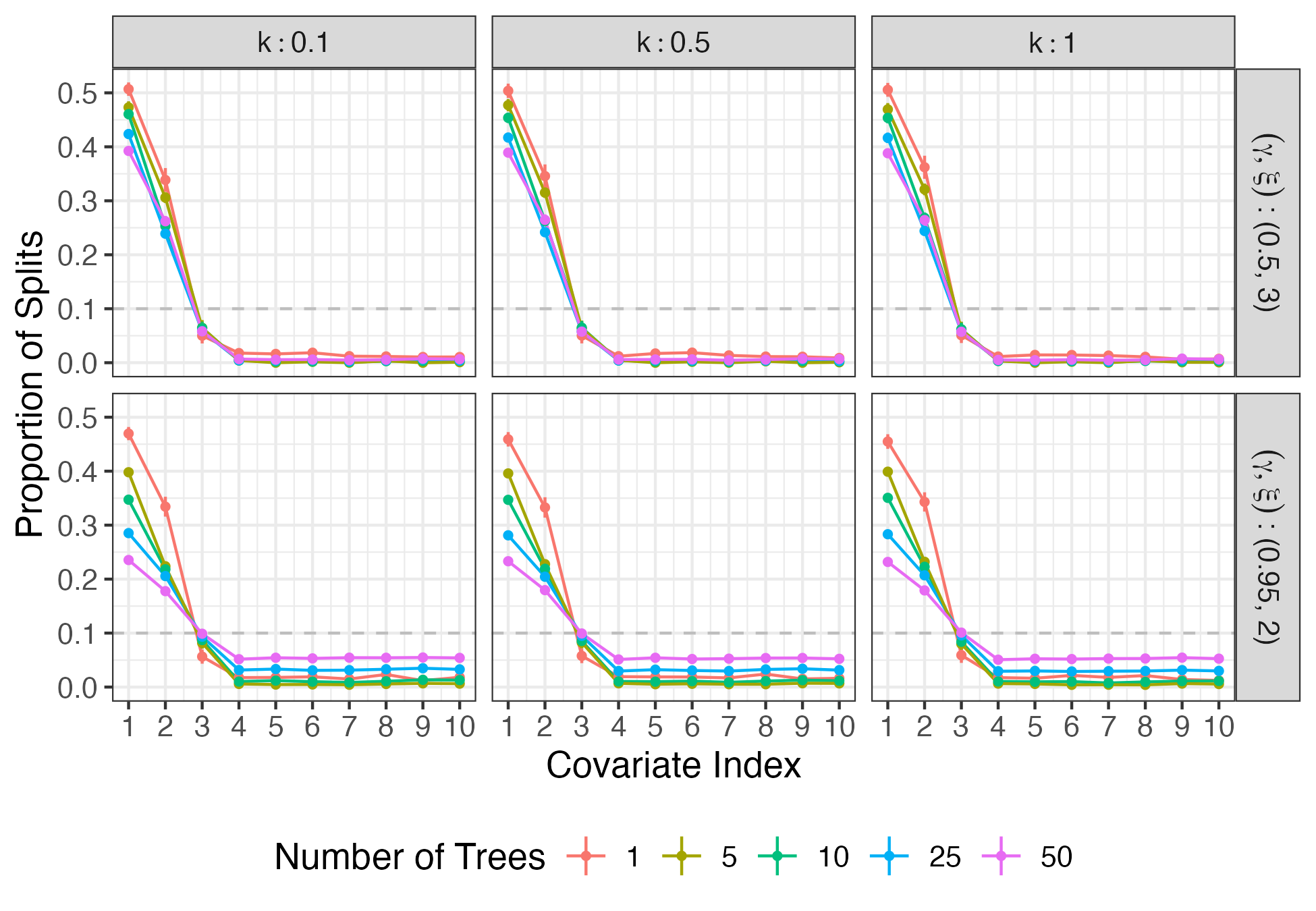}
    \caption{\textbf{Extended CART Simulation Variable Importance}: Plot of observed split proportions across 200 simulations (Monte Carlo mean and 95\% uncertainty interval presented).}
    \label{fig:ext-cart-sim-var-imp}
\end{figure}

\newpage
\begin{figure}[H]
    \centering
    \includegraphics[width=0.95\textwidth]{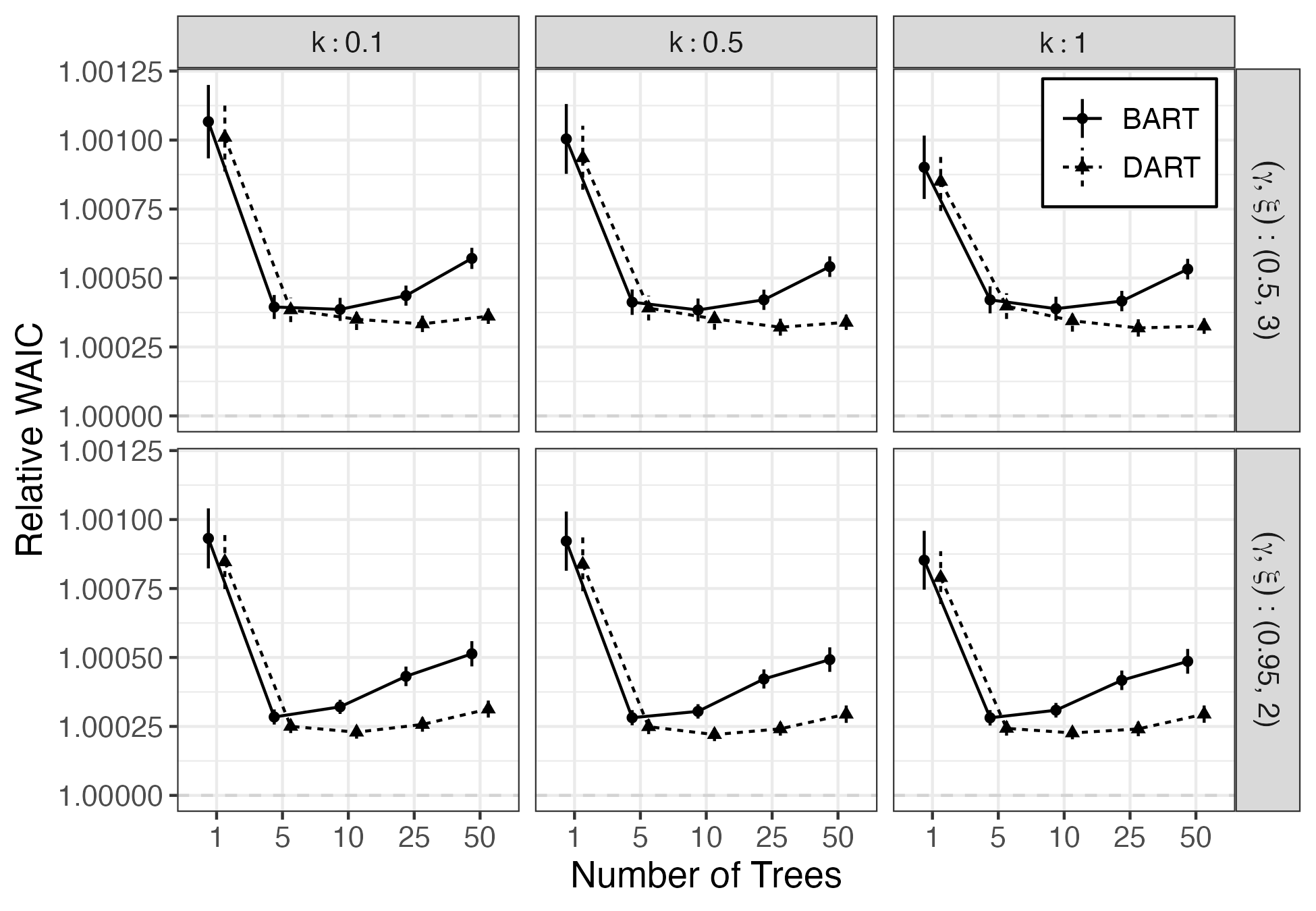}
    \caption{\textbf{WAIC for CART Simulation}: Relative WAIC for each simulation setting across 200 simulations (Monte Carlo mean and 95\% uncertainty intervals presented). BART included for reference to demonstrate the effect of the Dirichlet prior on covariate selection.}
    \label{fig:ext-cart-sim-waic}
\end{figure}

\newpage
\begin{figure}[H]
    \centering
    \includegraphics{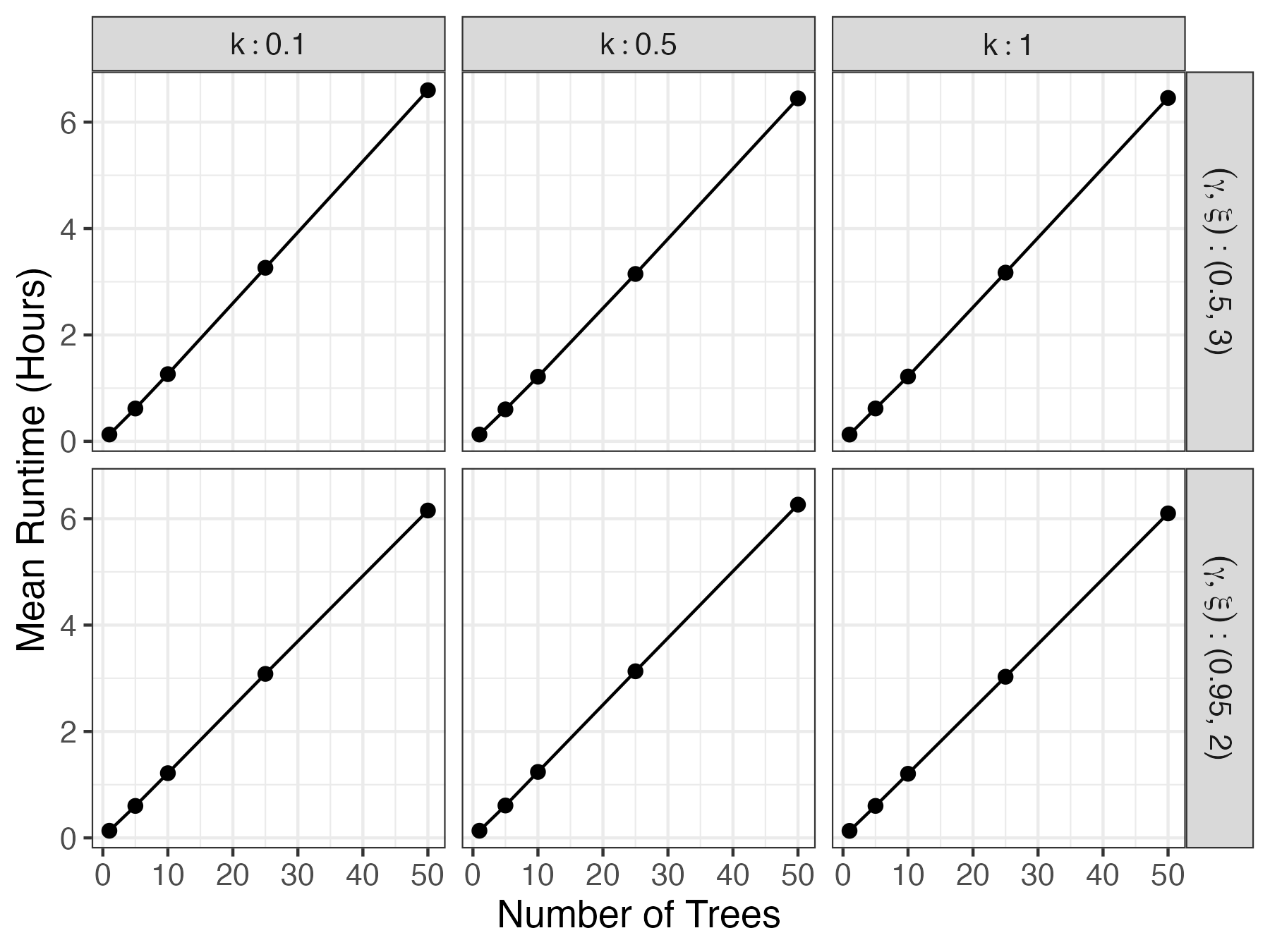}
    \caption{\textbf{CART Simulation Runtime}: Average time taken to run one CL-BART chain across 200 simulations. All models were run using 1 CPU on the high performance computing cluster at [redacted].}
    \label{fig:ext-cart-sim-time}
\end{figure}

\newpage
\spacingset{1.1}
\begin{table}[!h]

\caption{\label{tab:ext-friedman-sim-bart-stats}Extended Friedman Simulation Results - BART Predictions}
\centering
\fontsize{11}{13}\selectfont
\begin{tabular}[t]{cccccccc}
\toprule
Type & $M$\textsuperscript{a} & $k$\textsuperscript{b} & $(\gamma, \xi)$\textsuperscript{c} & Bias\textsuperscript{d} & RMSE\textsuperscript{d} & Coverage\textsuperscript{d} & Width\textsuperscript{d}\\
\midrule
oracle &  &  &  & -0.001 (0.001) & 0.040 (0.001) & 0.949 (0.016) & 0.160 (0.000)\\
\midrule
\addlinespace[2pt]
clbart & 5 & 0.1 & (0.5, 3) & -0.001 (0.001) & 0.169 (0.001) & 0.740 (0.031) & 0.387 (0.003)\\
clbart & 5 & 0.1 & (0.95, 2) & -0.001 (0.001) & 0.164 (0.001) & 0.804 (0.028) & 0.430 (0.003)\\
clbart & 5 & 0.5 & (0.5, 3) & -0.001 (0.001) & 0.170 (0.001) & 0.740 (0.031) & 0.390 (0.003)\\
clbart & 5 & 0.5 & (0.95, 2) & -0.001 (0.001) & 0.164 (0.001) & 0.807 (0.028) & 0.434 (0.003)\\
clbart & 5 & 1.0 & (0.5, 3) & -0.001 (0.001) & 0.171 (0.001) & 0.737 (0.031) & 0.388 (0.003)\\
clbart & 5 & 1.0 & (0.95, 2) & -0.001 (0.001) & 0.165 (0.001) & 0.801 (0.028) & 0.431 (0.002)\\
\midrule
\addlinespace[2pt]
clbart & 10 & 0.1 & (0.5, 3) & -0.001 (0.001) & 0.151 (0.001) & 0.846 (0.026) & 0.435 (0.002)\\
clbart & 10 & 0.1 & (0.95, 2) & -0.001 (0.001) & 0.142 (0.001) & 0.914 (0.020) & 0.498 (0.002)\\
clbart & 10 & 0.5 & (0.5, 3) & -0.001 (0.001) & 0.151 (0.001) & 0.845 (0.026) & 0.437 (0.002)\\
clbart & 10 & 0.5 & (0.95, 2) & -0.001 (0.001) & 0.143 (0.001) & 0.915 (0.020) & 0.502 (0.002)\\
clbart & 10 & 1.0 & (0.5, 3) & -0.001 (0.001) & 0.152 (0.001) & 0.842 (0.026) & 0.436 (0.002)\\
clbart & 10 & 1.0 & (0.95, 2) & -0.001 (0.001) & 0.144 (0.001) & 0.914 (0.020) & 0.502 (0.002)\\
\midrule
\addlinespace[2pt]
clbart & 25 & 0.1 & (0.5, 3) & -0.001 (0.001) & 0.138 (0.001) & 0.913 (0.020) & 0.477 (0.002)\\
clbart & 25 & 0.1 & (0.95, 2) & -0.001 (0.001) & 0.129 (0.001) & 0.967 (0.013) & 0.560 (0.002)\\
clbart & 25 & 0.5 & (0.5, 3) & -0.001 (0.001) & 0.138 (0.001) & 0.915 (0.020) & 0.479 (0.002)\\
clbart & 25 & 0.5 & (0.95, 2) & -0.001 (0.001) & 0.130 (0.001) & 0.967 (0.013) & 0.564 (0.002)\\
clbart & 25 & 1.0 & (0.5, 3) & -0.001 (0.001) & 0.138 (0.001) & 0.914 (0.020) & 0.480 (0.002)\\
clbart & 25 & 1.0 & (0.95, 2) & -0.001 (0.001) & 0.130 (0.001) & 0.967 (0.013) & 0.568 (0.002)\\
\midrule
\addlinespace[2pt]
clbart & 50 & 0.1 & (0.5, 3) & -0.001 (0.001) & 0.132 (0.001) & 0.942 (0.017) & 0.503 (0.002)\\
clbart & 50 & 0.1 & (0.95, 2) & -0.001 (0.001) & 0.126 (0.001) & 0.977 (0.011) & 0.583 (0.003)\\
clbart & 50 & 0.5 & (0.5, 3) & -0.001 (0.001) & 0.132 (0.001) & 0.943 (0.016) & 0.506 (0.002)\\
clbart & 50 & 0.5 & (0.95, 2) & -0.001 (0.001) & 0.126 (0.001) & 0.978 (0.010) & 0.590 (0.003)\\
clbart & 50 & 1.0 & (0.5, 3) & -0.001 (0.001) & 0.133 (0.001) & 0.943 (0.016) & 0.509 (0.002)\\
clbart & 50 & 1.0 & (0.95, 2) & -0.001 (0.001) & 0.127 (0.001) & 0.978 (0.010) & 0.596 (0.003)\\
\midrule
\addlinespace[2pt]
clbart & 100 & 0.1 & (0.5, 3) & -0.001 (0.001) & 0.128 (0.001) & 0.955 (0.015) & 0.520 (0.002)\\
clbart & 100 & 0.1 & (0.95, 2) & -0.001 (0.001) & 0.126 (0.001) & 0.979 (0.010) & 0.591 (0.003)\\
clbart & 100 & 0.5 & (0.5, 3) & -0.001 (0.001) & 0.129 (0.001) & 0.956 (0.014) & 0.523 (0.002)\\
clbart & 100 & 0.5 & (0.95, 2) & -0.001 (0.001) & 0.126 (0.001) & 0.980 (0.010) & 0.596 (0.003)\\
clbart & 100 & 1.0 & (0.5, 3) & -0.001 (0.001) & 0.129 (0.001) & 0.958 (0.014) & 0.529 (0.002)\\
clbart & 100 & 1.0 & (0.95, 2) & -0.001 (0.001) & 0.126 (0.001) & 0.980 (0.010) & 0.600 (0.003)\\
\bottomrule
\multicolumn{8}{l}{\rule{0pt}{1em}\textsuperscript{a} $M$: Number of trees.}\\
\multicolumn{8}{l}{\rule{0pt}{1em}\textsuperscript{b} $k$: Numerator of scale parameter for half-Cauchy hyper-prior.}\\
\multicolumn{8}{l}{\rule{0pt}{1em}\textsuperscript{c} $(\gamma, \xi)$: Hyperparameters for tree depth prior.}\\
\multicolumn{8}{l}{\rule{0pt}{1em}\textsuperscript{d} Monte Carlo mean and standard errors across 200 simulations reported.}\\
\end{tabular}
\end{table}
\spacingset{1.9}

\newpage
\spacingset{1.1}
\begin{table}[!h]

\caption{\label{tab:ext-friedman-sim-beta-bias}Extended Friedman Simulation Results - Confounder Estimates (Bias)}
\centering
\fontsize{11}{13}\selectfont
\begin{tabular}[t]{ccccccccc}
\toprule
\multicolumn{4}{c}{ } & \multicolumn{5}{c}{Bias $\times$ 1,000\textsuperscript{d}} \\
\cmidrule(l{3pt}r{3pt}){5-9}
Type & $M$\textsuperscript{a} & $k$\textsuperscript{b} & $(\gamma, \xi)$\textsuperscript{c} & $\beta_1$ & $\beta_2$ & $\beta_3$ & $\beta_4$ & $\beta_5$\\
\midrule
oracle &  &  &  & 2.28 (4.36) & -0.76 (4.37) & -4.09 (4.70) & 0.58 (4.57) & 4.06 (4.58)\\
\midrule
\addlinespace[2pt]
clbart & 5 & 0.1 & (0.5, 3) & -0.80 (4.37) & -1.52 (4.38) & -4.99 (4.73) & 1.60 (4.57) & 7.05 (4.63)\\
clbart & 5 & 0.1 & (0.95, 2) & -0.60 (4.35) & -2.09 (4.36) & -3.54 (4.72) & 1.59 (4.61) & 7.94 (4.60)\\
clbart & 5 & 0.5 & (0.5, 3) & -1.02 (4.37) & -1.24 (4.38) & -4.46 (4.72) & 1.28 (4.59) & 7.07 (4.57)\\
clbart & 5 & 0.5 & (0.95, 2) & -0.84 (4.34) & -1.05 (4.40) & -4.34 (4.70) & 1.66 (4.59) & 8.28 (4.58)\\
clbart & 5 & 1.0 & (0.5, 3) & -0.94 (4.32) & -1.18 (4.41) & -4.18 (4.71) & 1.29 (4.58) & 7.10 (4.58)\\
clbart & 5 & 1.0 & (0.95, 2) & -1.07 (4.34) & -1.51 (4.38) & -3.99 (4.70) & 1.60 (4.56) & 7.83 (4.59)\\
\midrule
\addlinespace[2pt]
clbart & 10 & 0.1 & (0.5, 3) & -0.22 (4.38) & -1.02 (4.39) & -4.29 (4.73) & 0.97 (4.55) & 7.36 (4.61)\\
clbart & 10 & 0.1 & (0.95, 2) & -1.04 (4.39) & -1.91 (4.36) & -4.32 (4.77) & 1.45 (4.65) & 7.76 (4.63)\\
clbart & 10 & 0.5 & (0.5, 3) & -0.10 (4.39) & -1.33 (4.39) & -4.51 (4.71) & 1.30 (4.58) & 7.27 (4.58)\\
clbart & 10 & 0.5 & (0.95, 2) & -1.14 (4.40) & -1.66 (4.42) & -3.49 (4.72) & 1.21 (4.60) & 8.18 (4.64)\\
clbart & 10 & 1.0 & (0.5, 3) & -0.80 (4.37) & -2.03 (4.42) & -4.25 (4.75) & 0.99 (4.60) & 7.55 (4.58)\\
clbart & 10 & 1.0 & (0.95, 2) & -0.48 (4.36) & -1.40 (4.41) & -4.42 (4.73) & 1.53 (4.58) & 8.01 (4.60)\\
\midrule
\addlinespace[2pt]
clbart & 25 & 0.1 & (0.5, 3) & -0.60 (4.38) & -1.47 (4.39) & -3.96 (4.71) & 1.32 (4.60) & 7.63 (4.61)\\
clbart & 25 & 0.1 & (0.95, 2) & -1.45 (4.39) & -1.53 (4.43) & -4.77 (4.71) & 1.15 (4.64) & 8.07 (4.61)\\
clbart & 25 & 0.5 & (0.5, 3) & -0.49 (4.38) & -1.26 (4.42) & -3.86 (4.71) & 1.56 (4.60) & 7.54 (4.59)\\
clbart & 25 & 0.5 & (0.95, 2) & -1.58 (4.36) & -2.28 (4.41) & -3.82 (4.72) & 1.86 (4.62) & 8.86 (4.61)\\
clbart & 25 & 1.0 & (0.5, 3) & -0.76 (4.41) & -1.95 (4.36) & -4.58 (4.72) & 1.59 (4.58) & 7.23 (4.62)\\
clbart & 25 & 1.0 & (0.95, 2) & -1.59 (4.39) & -1.94 (4.41) & -3.88 (4.74) & 1.81 (4.59) & 8.62 (4.60)\\
\midrule
\addlinespace[2pt]
clbart & 50 & 0.1 & (0.5, 3) & -0.23 (4.41) & -1.84 (4.36) & -3.92 (4.74) & 1.69 (4.60) & 7.60 (4.59)\\
clbart & 50 & 0.1 & (0.95, 2) & -1.72 (4.40) & -1.93 (4.37) & -3.87 (4.76) & 1.98 (4.62) & 8.86 (4.62)\\
clbart & 50 & 0.5 & (0.5, 3) & -0.42 (4.40) & -1.96 (4.42) & -4.27 (4.75) & 1.36 (4.63) & 7.88 (4.67)\\
clbart & 50 & 0.5 & (0.95, 2) & -1.77 (4.37) & -1.34 (4.43) & -4.08 (4.70) & 1.80 (4.63) & 9.19 (4.64)\\
clbart & 50 & 1.0 & (0.5, 3) & -1.09 (4.41) & -2.00 (4.41) & -3.96 (4.70) & 1.51 (4.60) & 7.98 (4.65)\\
clbart & 50 & 1.0 & (0.95, 2) & -1.91 (4.37) & -1.72 (4.41) & -4.13 (4.75) & 1.79 (4.64) & 8.58 (4.66)\\
\midrule
\addlinespace[2pt]
clbart & 100 & 0.1 & (0.5, 3) & -0.88 (4.38) & -2.29 (4.40) & -4.21 (4.72) & 0.85 (4.60) & 7.15 (4.60)\\
clbart & 100 & 0.1 & (0.95, 2) & -1.58 (4.39) & -1.89 (4.41) & -3.95 (4.75) & 1.34 (4.61) & 8.48 (4.66)\\
clbart & 100 & 0.5 & (0.5, 3) & -0.56 (4.40) & -1.71 (4.39) & -4.21 (4.73) & 1.48 (4.62) & 7.76 (4.59)\\
clbart & 100 & 0.5 & (0.95, 2) & -1.26 (4.37) & -1.50 (4.42) & -4.26 (4.73) & 1.55 (4.61) & 8.66 (4.62)\\
clbart & 100 & 1.0 & (0.5, 3) & -1.11 (4.39) & -1.81 (4.40) & -3.87 (4.69) & 1.20 (4.64) & 8.17 (4.64)\\
clbart & 100 & 1.0 & (0.95, 2) & -2.09 (4.41) & -1.98 (4.42) & -4.07 (4.73) & 2.08 (4.62) & 9.33 (4.62)\\
\bottomrule
\multicolumn{9}{l}{\rule{0pt}{1em}\textsuperscript{a} $M$: Number of trees.}\\
\multicolumn{9}{l}{\rule{0pt}{1em}\textsuperscript{b} $k$: Numerator of scale parameter for half-Cauchy hyper-prior.}\\
\multicolumn{9}{l}{\rule{0pt}{1em}\textsuperscript{c} $(\gamma, \xi)$: Hyperparameters for tree depth prior.}\\
\multicolumn{9}{l}{\rule{0pt}{1em}\textsuperscript{d} Monte Carlo mean and standard errors across 200 simulations reported.}\\
\end{tabular}
\end{table}
\spacingset{1.9}

\newpage
\spacingset{1.1}
\begin{table}[!h]

\caption{\label{tab:ext-friedman-sim-beta-coverage}Extended Friedman Simulation Results - Confounder Estimates (Coverage)}
\centering
\fontsize{11}{13}\selectfont
\begin{tabular}[t]{ccccccccc}
\toprule
\multicolumn{4}{c}{ } & \multicolumn{5}{c}{Coverage\textsuperscript{d}} \\
\cmidrule(l{3pt}r{3pt}){5-9}
Type & $M$\textsuperscript{a} & $k$\textsuperscript{b} & $(\gamma, \xi)$\textsuperscript{c} & $\beta_1$ & $\beta_2$ & $\beta_3$ & $\beta_4$ & $\beta_5$\\
\midrule
oracle &  &  &  & 0.98 & 0.96 & 0.92 & 0.95 & 0.96\\
\midrule
\addlinespace[2pt]
clbart & 5 & 0.1 & (0.5, 3) & 0.97 & 0.96 & 0.92 & 0.93 & 0.94\\
clbart & 5 & 0.1 & (0.95, 2) & 0.98 & 0.96 & 0.93 & 0.93 & 0.96\\
clbart & 5 & 0.5 & (0.5, 3) & 0.97 & 0.97 & 0.93 & 0.93 & 0.95\\
clbart & 5 & 0.5 & (0.95, 2) & 0.97 & 0.96 & 0.94 & 0.95 & 0.94\\
clbart & 5 & 1.0 & (0.5, 3) & 0.98 & 0.96 & 0.93 & 0.93 & 0.94\\
clbart & 5 & 1.0 & (0.95, 2) & 0.98 & 0.95 & 0.92 & 0.94 & 0.96\\
\midrule
\addlinespace[2pt]
clbart & 10 & 0.1 & (0.5, 3) & 0.98 & 0.96 & 0.92 & 0.95 & 0.95\\
clbart & 10 & 0.1 & (0.95, 2) & 0.97 & 0.96 & 0.92 & 0.93 & 0.95\\
clbart & 10 & 0.5 & (0.5, 3) & 0.97 & 0.96 & 0.92 & 0.93 & 0.95\\
clbart & 10 & 0.5 & (0.95, 2) & 0.97 & 0.95 & 0.92 & 0.93 & 0.94\\
clbart & 10 & 1.0 & (0.5, 3) & 0.97 & 0.96 & 0.94 & 0.93 & 0.95\\
clbart & 10 & 1.0 & (0.95, 2) & 0.97 & 0.95 & 0.92 & 0.93 & 0.95\\
\midrule
\addlinespace[2pt]
clbart & 25 & 0.1 & (0.5, 3) & 0.98 & 0.97 & 0.92 & 0.93 & 0.96\\
clbart & 25 & 0.1 & (0.95, 2) & 0.97 & 0.96 & 0.92 & 0.94 & 0.94\\
clbart & 25 & 0.5 & (0.5, 3) & 0.97 & 0.97 & 0.92 & 0.93 & 0.95\\
clbart & 25 & 0.5 & (0.95, 2) & 0.97 & 0.96 & 0.93 & 0.94 & 0.95\\
clbart & 25 & 1.0 & (0.5, 3) & 0.97 & 0.96 & 0.92 & 0.93 & 0.95\\
clbart & 25 & 1.0 & (0.95, 2) & 0.98 & 0.96 & 0.92 & 0.92 & 0.95\\
\midrule
\addlinespace[2pt]
clbart & 50 & 0.1 & (0.5, 3) & 0.96 & 0.96 & 0.92 & 0.94 & 0.95\\
clbart & 50 & 0.1 & (0.95, 2) & 0.97 & 0.96 & 0.93 & 0.94 & 0.95\\
clbart & 50 & 0.5 & (0.5, 3) & 0.97 & 0.96 & 0.92 & 0.94 & 0.95\\
clbart & 50 & 0.5 & (0.95, 2) & 0.97 & 0.96 & 0.92 & 0.93 & 0.95\\
clbart & 50 & 1.0 & (0.5, 3) & 0.97 & 0.95 & 0.93 & 0.94 & 0.94\\
clbart & 50 & 1.0 & (0.95, 2) & 0.97 & 0.96 & 0.92 & 0.92 & 0.96\\
\midrule
\addlinespace[2pt]
clbart & 100 & 0.1 & (0.5, 3) & 0.97 & 0.96 & 0.93 & 0.94 & 0.95\\
clbart & 100 & 0.1 & (0.95, 2) & 0.97 & 0.96 & 0.92 & 0.95 & 0.96\\
clbart & 100 & 0.5 & (0.5, 3) & 0.96 & 0.95 & 0.92 & 0.93 & 0.95\\
clbart & 100 & 0.5 & (0.95, 2) & 0.97 & 0.95 & 0.94 & 0.93 & 0.95\\
clbart & 100 & 1.0 & (0.5, 3) & 0.97 & 0.96 & 0.92 & 0.94 & 0.94\\
clbart & 100 & 1.0 & (0.95, 2) & 0.97 & 0.96 & 0.93 & 0.93 & 0.96\\
\bottomrule
\multicolumn{9}{l}{\rule{0pt}{1em}\textsuperscript{a} $M$: Number of trees.}\\
\multicolumn{9}{l}{\rule{0pt}{1em}\textsuperscript{b} $k$: Numerator of scale parameter for half-Cauchy hyper-prior.}\\
\multicolumn{9}{l}{\rule{0pt}{1em}\textsuperscript{c} $(\gamma, \xi)$: Hyperparameters for tree depth prior.}\\
\multicolumn{9}{l}{\rule{0pt}{1em}\textsuperscript{d} 95\% credible interval coverage across 200 simulations.}\\
\end{tabular}
\end{table}
\spacingset{1.9}

\begin{figure}[H]
    \centering
    \includegraphics[width=0.95\textwidth]{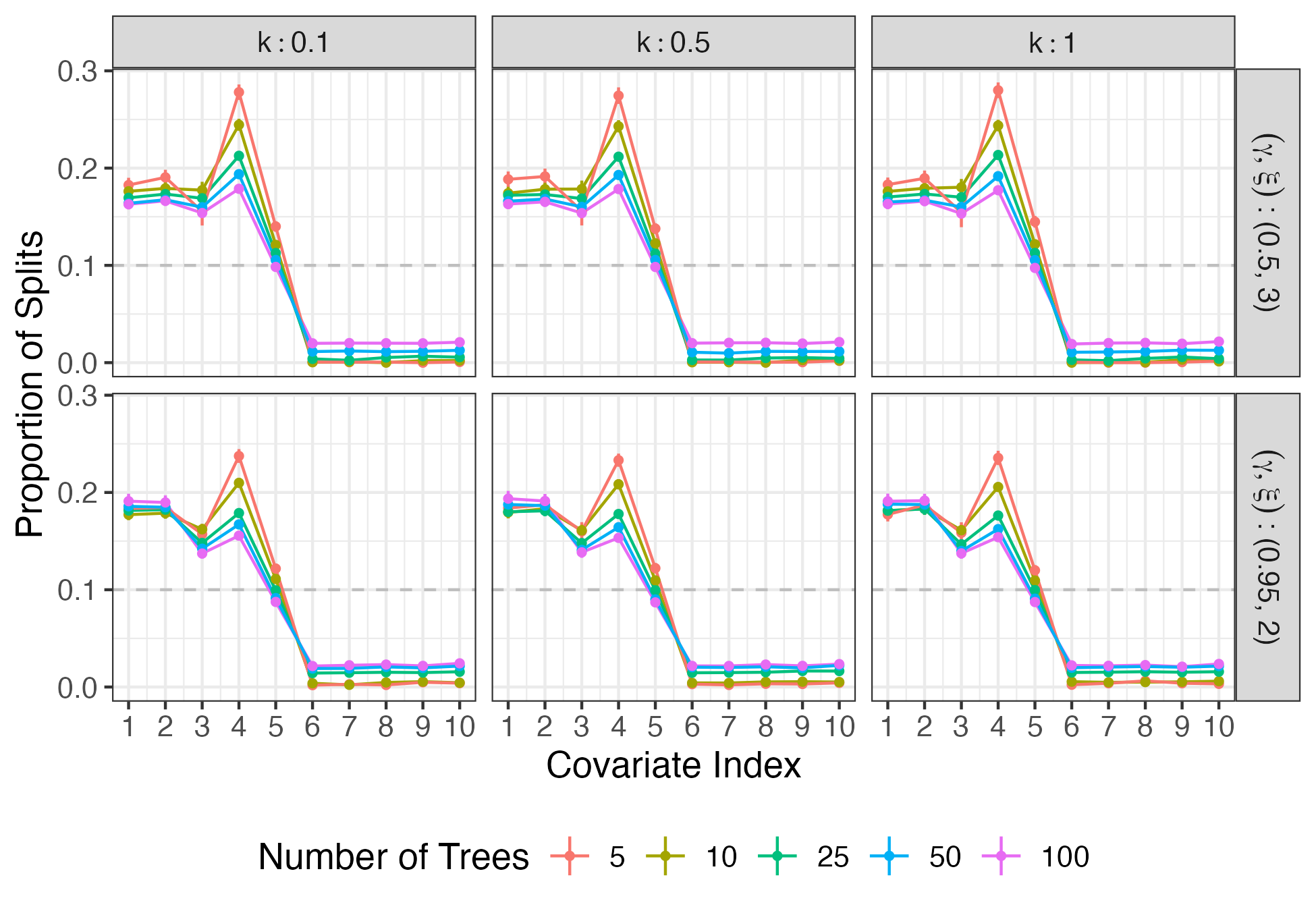}
    \caption{\textbf{Extended Friedman Simulation Variable Importance}: Plot of observed split proportions across 200 simulations (Monte Carlo mean and 95\% uncertainty interval presented).}
    \label{fig:ext-friedman-sim-var-imp}
\end{figure}

\begin{figure}[H]
    \centering
    \includegraphics[width=0.95\textwidth]{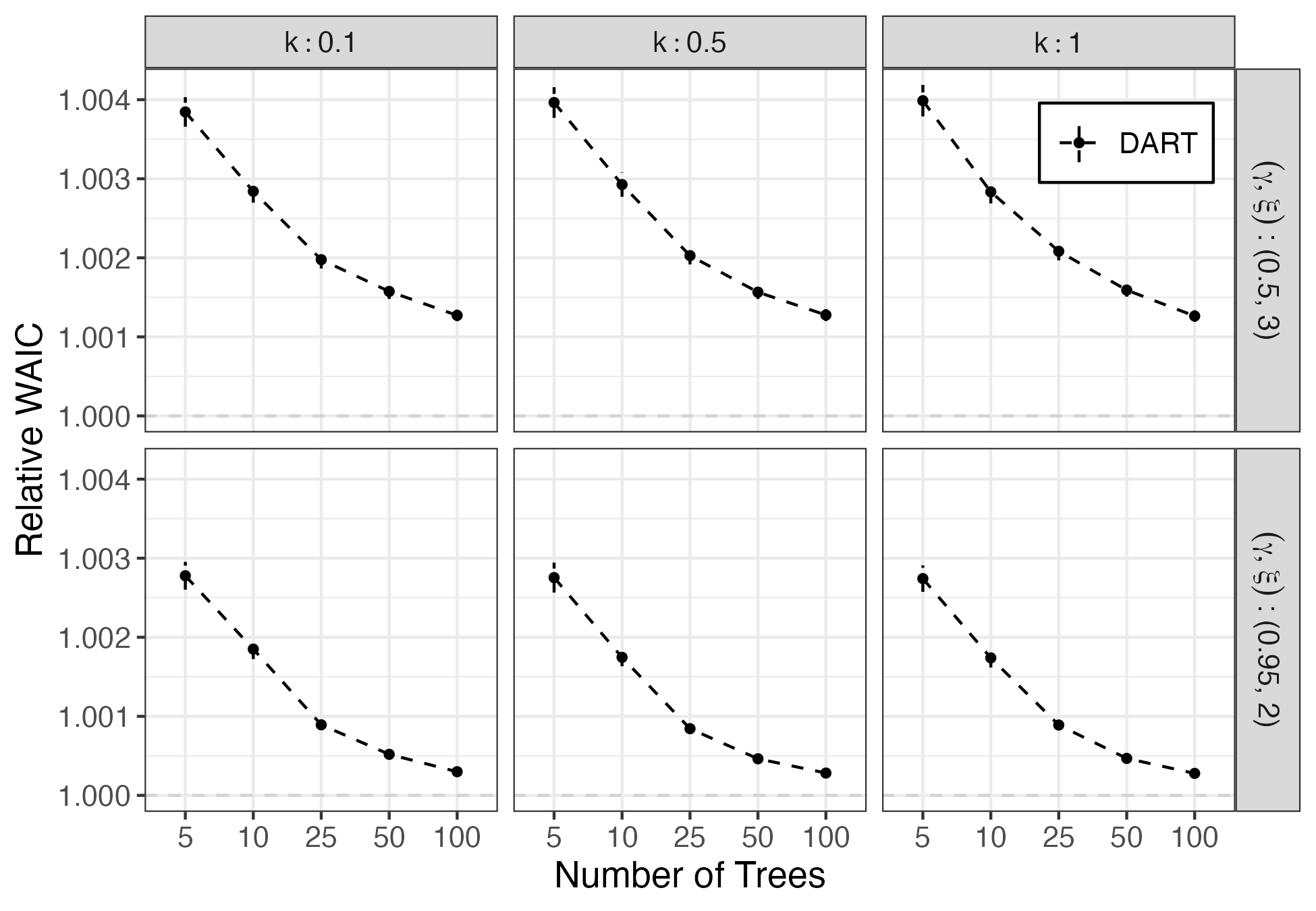}
    \caption{\textbf{WAIC for Friedman Simulation}: Relative WAIC for each simulation setting across 200 simulations (Monte Carlo mean and 95\% uncertainty intervals presented).}
    \label{fig:ext-friedman-sim-waic}
\end{figure}

\begin{figure}[H]
    \centering
    \includegraphics[width=0.95\textwidth]{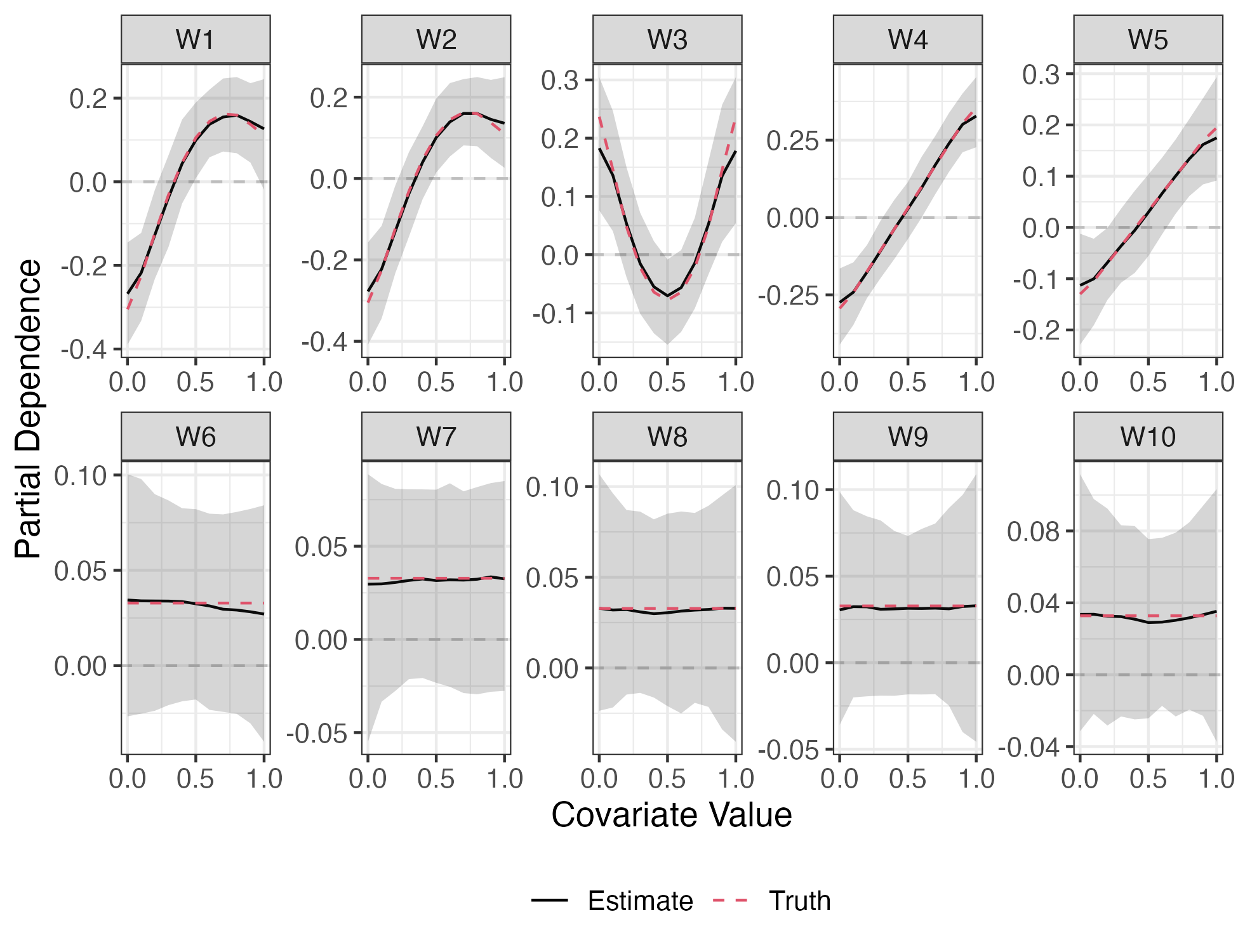}
    \caption{\textbf{Partial Dependence Plots for Friedman Simulation}: Estimates of the marginal partial dependence function for each covariate across 200 simulations with $(M,k,\gamma, \xi) = (100,1,0.95,2)$ (simulation mean and 95\% quantile intervals presented).}
    \label{fig:friedman-sim-pdp}
\end{figure}

\begin{figure}[H]
    \centering
    \includegraphics{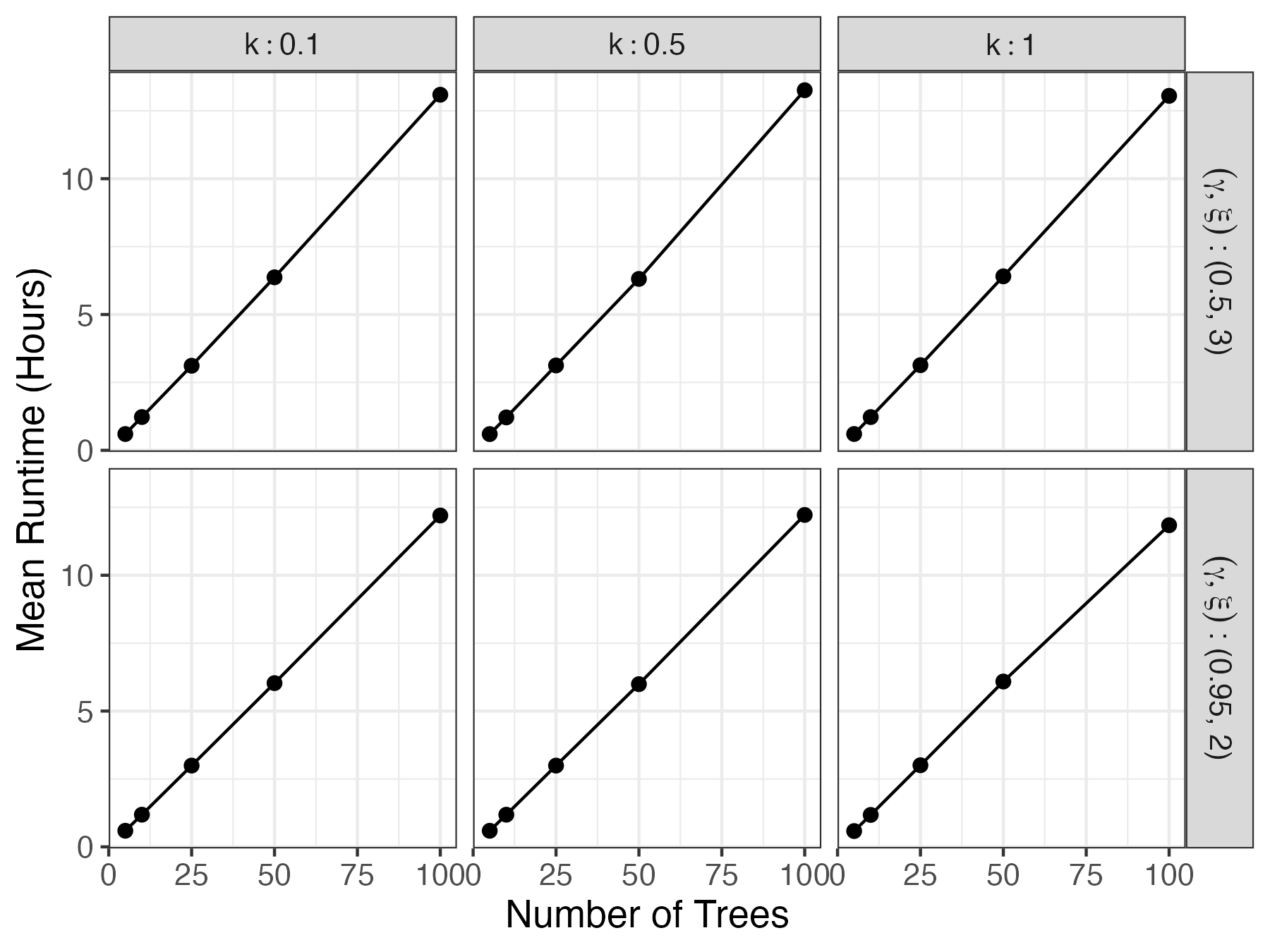}
    \caption{\textbf{Friedman Simulation Runtime}: Average time taken to run one CL-BART chain across 200 simulations. All models were run using 1 CPU on the high performance computing cluster at [redacted].}
    \label{fig:ext-friedman-sim-time}
\end{figure}

\newpage

\section{Additional Case Study Materials}
\begin{figure}[H]
    \centering
    \includegraphics[width=0.8\textwidth]{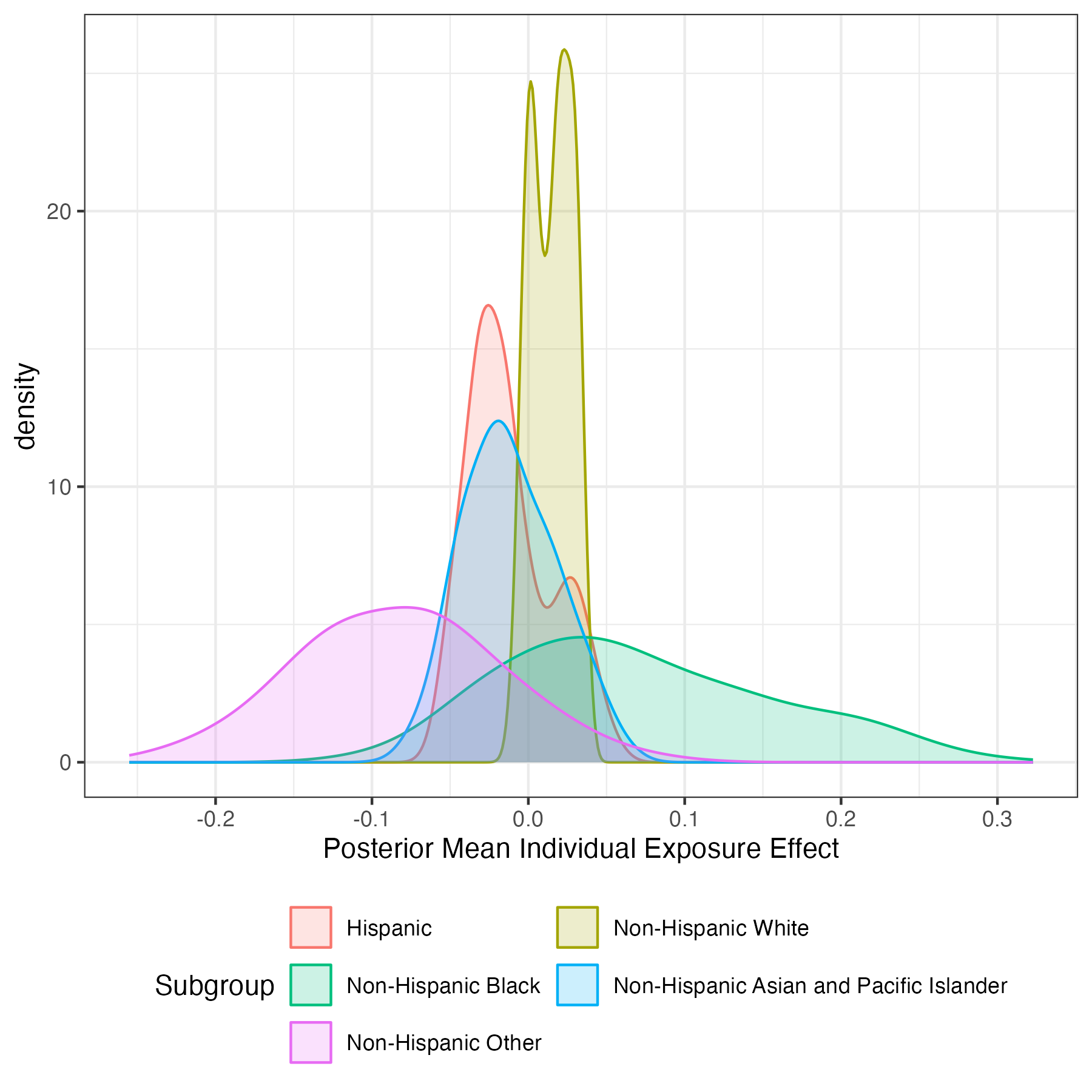}
    \caption{\textbf{Distribution of Individual Exposure Effects}: Density plots demonstrating the heterogeneity in point estimates of individual conditional exposure effects.}
    \label{fig:app-preds-density}
\end{figure}

\spacingset{1.1}
\begin{landscape}\begin{table}

\caption{\label{tab:app-subgroup-descriptives}Descriptive Statistics for AD ED Patients by Subgroup, CA 2005-2015}
\centering
\fontsize{12}{14}\selectfont
\begin{tabular}[t]{lrrrrr}
\toprule
Characteristic & HISP & NHW & NHB & NHAPI & NHO\\
\midrule
n & 11,959 & 46,019 & 5,635 & 5,521 & 1,886\\
\addlinespace[2pt]
Sex &  &  &  &  & \\
\hspace{1em}Male & 4,356 (36.4\%) & 16,857 (36.6\%) & 1,839 (32.6\%) & 1,979 (35.8\%) & 731 (38.8\%)\\
\hspace{1em}Female & 7,603 (63.6\%) & 29,162 (63.4\%) & 3,796 (67.4\%) & 3,542 (64.2\%) & 1,155 (61.2\%)\\
\addlinespace[2pt]
Age, yrs & 83 (78, 88) & 85 (80, 89) & 82 (77, 88) & 85 (80, 89) & 84 (78, 88)\\
\addlinespace[2pt]
Number of Comorbid Conditions & 2 (1, 3) & 2 (1, 3) & 2 (1, 3) & 2 (1, 3) & 2 (1, 3)\\
\addlinespace[2pt]
CHF & 1,965 (16.4\%) & 7,287 (15.8\%) & 965 (17.1\%) & 960 (17.4\%) & 317 (16.8\%)\\
\addlinespace[2pt]
CKD & 3,042 (25.4\%) & 10,775 (23.4\%) & 1,925 (34.2\%) & 1,784 (32.3\%) & 411 (21.8\%)\\
\addlinespace[2pt]
COPD & 1,319 (11.0\%) & 5,655 (12.3\%) & 644 (11.4\%) & 670 (12.1\%) & 195 (10.3\%)\\
\addlinespace[2pt]
Depression & 1,519 (12.7\%) & 6,259 (13.6\%) & 432 (7.7\%) & 546 (9.9\%) & 249 (13.2\%)\\
\addlinespace[2pt]
Diabetes & 4,605 (38.5\%) & 8,687 (18.9\%) & 1,788 (31.7\%) & 2,066 (37.4\%) & 508 (26.9\%)\\
\addlinespace[2pt]
Hypertension & 8,212 (68.7\%) & 28,500 (61.9\%) & 4,296 (76.2\%) & 4,085 (74.0\%) & 1,188 (63.0\%)\\
\addlinespace[2pt]
Hyperlipidemia & 3,610 (30.2\%) & 13,731 (29.8\%) & 1,766 (31.3\%) & 1,936 (35.1\%) & 532 (28.2\%)\\
\addlinespace[2pt]
\bottomrule
\multicolumn{6}{l}{\rule{0pt}{1em}n: sample size.}\\
\multicolumn{6}{l}{\rule{0pt}{1em}Median (IQR) reported for age. number of conditions.}\\
\multicolumn{6}{l}{\rule{0pt}{1em}HISP: Hispanic, NHW: Non-Hispanic White, NHB: Non-Hispanic Black, NHAPI: Non-Hispanic Asian and }\\
\multicolumn{6}{l}{\rule{0pt}{1em}Pacific Islander, NHO: Non-Hispanic Other.}\\
\multicolumn{6}{l}{\rule{0pt}{1em}CHF: Congestive Heart Failure, CKD: Chronic Kidney Disease, COPD: Chronic Obstructive Pulmonary Disease.}\\
\end{tabular}
\end{table}
\end{landscape}
\spacingset{1.9}

\spacingset{1.1}
\begin{table}

\caption{\label{tab:app-marg-pd-tbl}Marginal Partial Dependence Estimates}
\centering
\fontsize{11}{12}\selectfont
\begin{tabular}[t]{llc}
\toprule
Subgroup & Covariate & $\exp{\left( \bar{\tau}_{diff} \right)}^a$\\
\midrule
 & Female & 0.99 (0.93, 1.02)\\

 & CHF & 0.99 (0.92, 1.04)\\

 & CKD & 1.06 (0.99, 1.20)\\

 & COPD & 1.00 (0.94, 1.07)\\

 & Depression & 1.01 (0.95, 1.08)\\

 & Diabetes & 0.98 (0.92, 1.02)\\

 & Hypertension & 0.99 (0.93, 1.02)\\

\multirow{-8}{*}{\raggedright\arraybackslash Hispanic} & Hyperlipidemia & 1.00 (0.96, 1.05)\\
\cmidrule{1-3}
 & Female & 1.00 (0.99, 1.02)\\

 & CHF & 1.00 (0.96, 1.01)\\

 & CKD & 1.01 (0.99, 1.04)\\

 & COPD & 1.00 (0.96, 1.02)\\

 & Depression & 0.99 (0.96, 1.02)\\

 & Diabetes & 1.00 (0.97, 1.02)\\

 & Hypertension & 1.00 (0.98, 1.01)\\

\multirow{-8}{*}{\raggedright\arraybackslash Non-Hispanic White} & Hyperlipidemia & 1.01 (0.99, 1.04)\\
\cmidrule{1-3}
 & Female & 1.01 (0.93, 1.10)\\

 & CHF & 1.01 (0.91, 1.13)\\

 & CKD & 1.05 (0.96, 1.17)\\

 & COPD & 0.93 (0.79, 1.04)\\

 & Depression & 0.97 (0.81, 1.11)\\

 & Diabetes & 1.06 (0.98, 1.19)\\

 & Hypertension & 0.85 (0.74, 1.00)\\

\multirow{-8}{*}{\raggedright\arraybackslash Non-Hispanic Black} & Hyperlipidemia & 1.01 (0.94, 1.12)\\
\cmidrule{1-3}
 & Female & 1.02 (0.97, 1.11)\\

 & CHF & 1.03 (0.96, 1.15)\\

 & CKD & 1.02 (0.96, 1.13)\\

 & COPD & 1.00 (0.91, 1.09)\\

 & Depression & 0.99 (0.87, 1.06)\\

 & Diabetes & 1.04 (0.98, 1.13)\\

 & Hypertension & 1.01 (0.94, 1.08)\\

\multirow{-8}{*}{\raggedright\arraybackslash Non-Hispanic Asian and Pacific Islander} & Hyperlipidemia & 1.01 (0.95, 1.07)\\
\cmidrule{1-3}
 & Female & 0.94 (0.78, 1.04)\\

 & CHF & 1.02 (0.90, 1.22)\\

 & CKD & 1.09 (0.93, 1.35)\\

 & COPD & 1.00 (0.84, 1.19)\\

 & Depression & 1.03 (0.89, 1.22)\\

 & Diabetes & 0.96 (0.79, 1.05)\\

 & Hypertension & 1.00 (0.91, 1.14)\\

\multirow{-8}{*}{\raggedright\arraybackslash Non-Hispanic Other} & Hyperlipidemia & 0.96 (0.82, 1.05)\\
\bottomrule
\multicolumn{3}{l}{\rule{0pt}{1em}\textsuperscript{a} $\bar{\tau}_{diff}$: Difference in marginal partial average exposure effects.}\\
\multicolumn{3}{l}{\rule{0pt}{1em}Posterior mean and 95\% credible interval presented.}\\
\end{tabular}
\end{table}
\spacingset{1.9}

\begin{figure}[H]
    \centering
    \includegraphics[width = \textwidth]{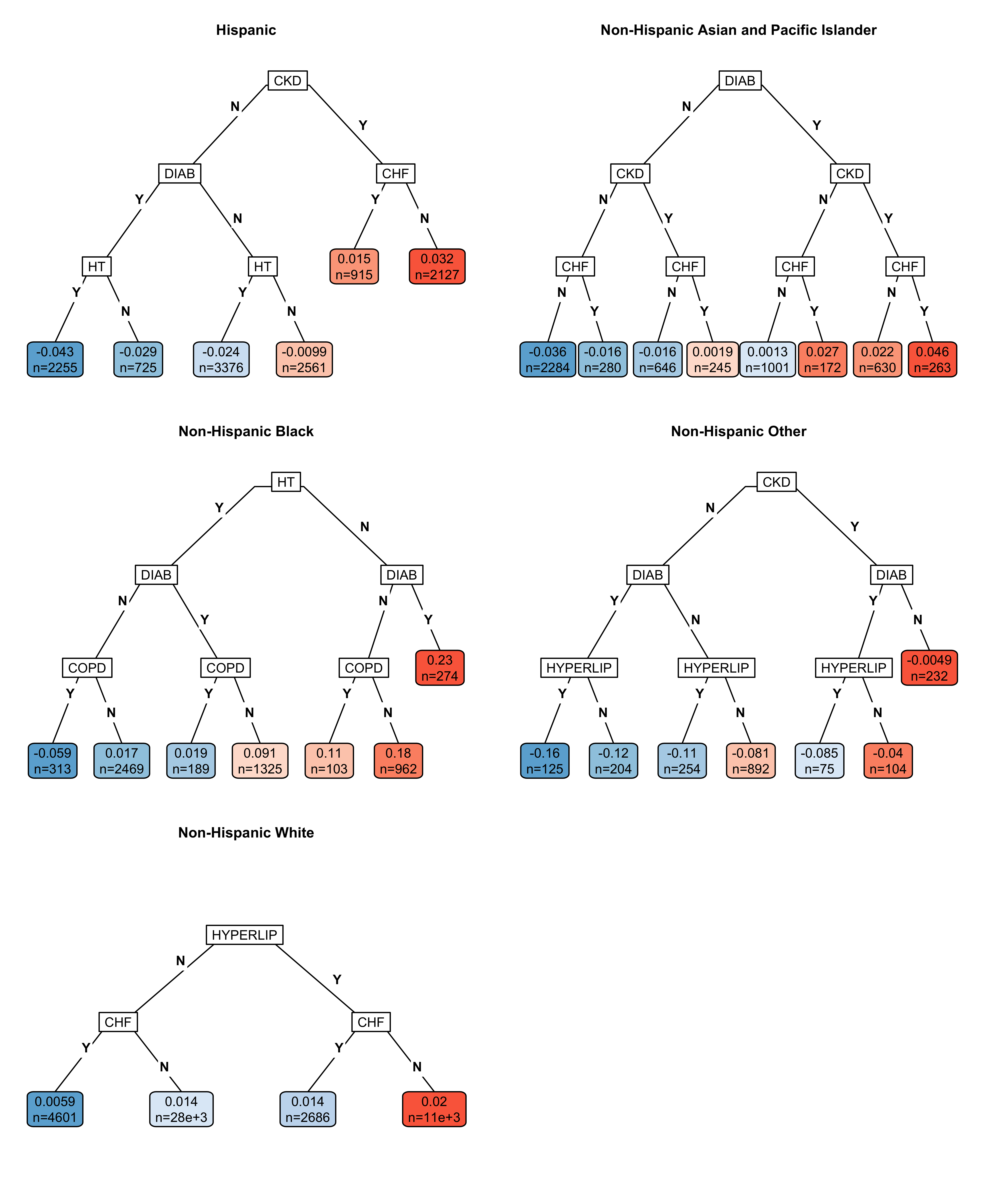}
    \vspace{-3em}
    \caption{\textbf{Lower-dimensional CART summaries}: CART diagrams of lower-dimensional summaries for CL-BART model predictions (log odds ratio scale).}
    \label{fig:app-cart-summaries}
\end{figure}

\spacingset{1.1}
\begin{table}

\caption{\label{tab:app-cart-pd-tbl}Lower Dimensional CART Summary Partial Dependence Estimates}
\centering
\fontsize{11}{12}\selectfont
\begin{tabular}[t]{cccccccccc}
\toprule
CHF & CKD & COPD & DEP & DIAB & HT & HLD & $\exp{\left( \bar{\tau} \right)}^a$ & $\Pr\left(\bar{\tau} > 0 \right)$ & $R^2$ $^b$\\
\midrule
\addlinespace[0.3em]
\multicolumn{10}{l}{\textbf{Hispanic}}\\
\hline
+ & + &  &  &  &  &  & 1.02 (0.94, 1.13) & 0.65 & 0.89\\
- & + &  &  &  &  &  & 1.04 (0.96, 1.17) & 0.74 & 0.89\\
 & - &  &  & + & + &  & 0.96 (0.85, 1.03) & 0.20 & 0.89\\
 & - &  &  & + & - &  & 0.97 (0.88, 1.04) & 0.27 & 0.89\\
 & - &  &  & - & + &  & 0.98 (0.90, 1.04) & 0.28 & 0.89\\
 & - &  &  & - & - &  & 0.99 (0.92, 1.06) & 0.40 & 0.89\\
\addlinespace[0.3em]
\hline
\multicolumn{10}{l}{\textbf{Non-Hispanic White}}\\
\hline
+ &  &  &  &  &  & - & 1.01 (0.97, 1.05) & 0.69 & 0.09\\
- &  &  &  &  &  & - & 1.01 (0.98, 1.05) & 0.79 & 0.09\\
+ &  &  &  &  &  & + & 1.02 (0.98, 1.06) & 0.75 & 0.09\\
- &  &  &  &  &  & + & 1.02 (0.99, 1.07) & 0.84 & 0.09\\
\addlinespace[0.3em]
\hline
\multicolumn{10}{l}{\textbf{Non-Hispanic Black}}\\
\hline
 &  &  &  & + & - &  & 1.27 (1.04, 1.53) & 0.99 & 0.79\\
 &  & + &  & - & + &  & 0.95 (0.77, 1.11) & 0.29 & 0.79\\
 &  & - &  & - & + &  & 1.02 (0.90, 1.17) & 0.63 & 0.79\\
 &  & + &  & + & + &  & 1.02 (0.84, 1.20) & 0.56 & 0.79\\
 &  & - &  & + & + &  & 1.09 (0.96, 1.25) & 0.88 & 0.79\\
 &  & + &  & - & - &  & 1.13 (0.92, 1.39) & 0.89 & 0.79\\
 &  & - &  & - & - &  & 1.22 (1.02, 1.43) & 0.99 & 0.79\\
\addlinespace[0.3em]
\hline
\multicolumn{10}{l}{\textbf{Non-Hispanic Asian and Pacific Islander}}\\
\hline
- & - &  &  & - &  &  & 0.97 (0.86, 1.07) & 0.25 & 0.71\\
+ & - &  &  & - &  &  & 0.99 (0.88, 1.13) & 0.40 & 0.71\\
- & + &  &  & - &  &  & 0.99 (0.89, 1.11) & 0.41 & 0.71\\
+ & + &  &  & - &  &  & 1.01 (0.90, 1.16) & 0.53 & 0.71\\
- & - &  &  & + &  &  & 1.00 (0.89, 1.12) & 0.48 & 0.71\\
+ & - &  &  & + &  &  & 1.03 (0.91, 1.21) & 0.65 & 0.71\\
- & + &  &  & + &  &  & 1.02 (0.91, 1.16) & 0.62 & 0.71\\
+ & + &  &  & + &  &  & 1.05 (0.93, 1.23) & 0.74 & 0.71\\
\addlinespace[0.3em]
\hline
\multicolumn{10}{l}{\textbf{Non-Hispanic Other}}\\
\hline
 & + &  &  & - &  &  & 1.01 (0.80, 1.27) & 0.46 & 0.39\\
 & - &  &  & + &  & + & 0.86 (0.63, 1.05) & 0.10 & 0.39\\
 & - &  &  & + &  & - & 0.90 (0.69, 1.08) & 0.16 & 0.39\\
 & - &  &  & - &  & + & 0.90 (0.71, 1.09) & 0.16 & 0.39\\
 & - &  &  & - &  & - & 0.94 (0.76, 1.11) & 0.24 & 0.39\\
 & + &  &  & + &  & + & 0.93 (0.72, 1.17) & 0.25 & 0.39\\
 & + &  &  & + &  & - & 0.97 (0.76, 1.19) & 0.34 & 0.39\\
\bottomrule
\multicolumn{10}{l}{\rule{0pt}{1em}\textsuperscript{a} $\bar{\tau}$: Partial average exposure effect. Posterior mean and 95\% credible interval presented.}\\
\multicolumn{10}{l}{\rule{0pt}{1em}\textsuperscript{b} Summary $R^2$.}\\
\end{tabular}
\end{table}
\spacingset{1.9}

\begin{figure}[H]
    \centering
    \includegraphics[width = \textwidth]{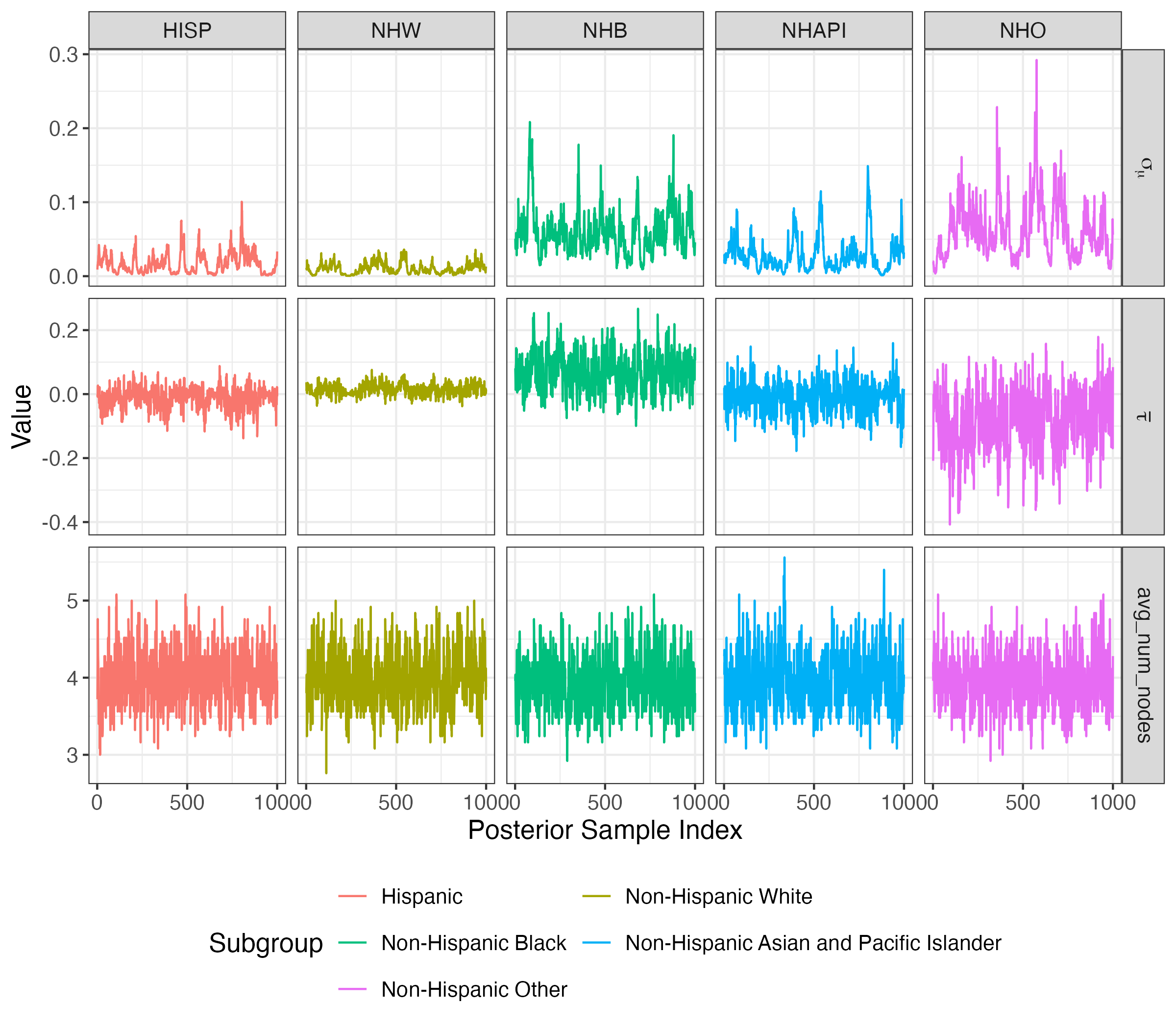}
    \caption{\textbf{Trace Plots for Selected Parameters}: Trace plots for each subgroup. Parameters from top to bottom include: $\sigma_\mu$ (leaf node prior standard deviation), $\bar{\tau}$ (average exposure effect, on the log scale), and the average number of nodes across all 25 trees.}
    \label{fig:app-diag-trace}
\end{figure}

\spacingset{1.1}
\begin{table}

\caption{\label{tab:app-timing}Application Model Runtime}
\centering
\fontsize{12}{14}\selectfont
\begin{threeparttable}
\begin{tabular}[t]{lrr}
\toprule
Subgroup & Sample Size & Runtime$^a$\\
\midrule
Hispanic & 11959 & 6.28\\
Non-Hispanic White & 46019 & 23.67\\
Non-Hispanic Black & 5635 & 3.69\\
Non-Hispanic Asian and Pacific Islander & 5521 & 3.72\\
Non-Hispanic Other & 1886 & 1.26\\
\bottomrule
\end{tabular}
\begin{tablenotes}
\item[a] Time taken to run one chain of a 25-tree CL-BART model, in hours. All models were run using 1 CPU on the high performance computing cluster at [redacted].
\end{tablenotes}
\end{threeparttable}
\end{table}
\spacingset{1.9}

\newpage

\bibliographystyle{apalike}
\bibliography{references}